\documentclass[twocolumn,tighten]{aastex61}
\usepackage{sidecap}

\begin{document}

\def\degr{\ensuremath{^\circ}}
\def\arcmin{\ensuremath{^\prime}}
\def\arcsec{\ensuremath{^{\prime\prime}}}
\def\Kkms{\rm K\,km\,s^{-1}}
\def\kms{\rm km\,s^{-1}}
\def\ms{\rm m\,s^{-1}}
\def\percc{\rm cm^{-3}}
\def\cms{\rm cm\,s^{-1}}
\def\pers{\rm s^{-1}}
\def\mum{\rm \mu m}
\def\persqcm{\rm cm^{-2}}
\def\sqcm{\rm cm^2}

\def\utw{\smash{\rlap{\lower5pt\hbox{$\sim$}}}}
\def\udtw{\smash{\rlap{\lower6pt\hbox{$\approx$}}}}
\def\fd{\hbox{$.\!\!^{\rm d}$}} \def\fh{\hbox{$.\!\!^{\rm h}$}}
\def\fm{\hbox{$.\!\!^{\rm m}$}} \def\fs{\hbox{$.\!\!^{\rm s}$}}
\def\fdg{\hbox{$.\!\!^\circ$}} \def\farcm{\hbox{$.\mkern-4mu^\prime$}}
\def\farcs{\hbox{$.\!\!^{\prime\prime}$}} 

\def\ammo{\rm NH_3}
\def\dammo{\rm NH_2D} \def\ddammo{\rm NHD_2} \def\ndthree{\rm ND_3}
\def\diaz{\rm N_2H^+} \def\ddiaz{\rm N_2D^+} \def\hthree{\rm H_3^+}
\def\htwod{\rm H_2D^+} \def\dtwoh{\rm D_2H^+} \def\dthree{\rm D_3^+}
\def\htwo{\rm H_2} 
\def\dtwo{\rm D_2} 
\def\Harpoons{\mathop{\rightleftharpoons}\limits}
\def\el{\rm e^-}
\def\arrow{\mathop{\rightarrow}\limits}
\def\meth{\rm CH_3OH}
\def\forma{\rm H_2CO}
\def\water{\rm H_2O}
\def\hone{\rm H}
\def\hd{\rm HD}
\def\formyl{\rm HCO}
\def\methoxy{\rm CH_3O}
\def\hydroxymethyl{\rm CH_2OH}
\def\co{\rm CO}
\def\cotwo{\rm CO_2}
\def\methane{\rm CH_4}

\def\dimethether{\rm CH_3OCH_3}
\def\methform{\rm HCOOCH_3}
\def\acetonitrile{\rm CH_3CN}
\def\methacet{\rm CH_3CCH}
\def\sio{\rm SiO}
\def\otwo{\rm O_2}
\def\sotwo{\rm SO_2}
\def\htwos{\rm H_2S} 

%\received{...}
%\revised{...}
%\accepted{...}
%\submitjournal{ApJL}

%\shorttitle{\aastex\ sample article}
%\shortauthors{Schwarz et al.}

\title{Efficient methanol production on the dark side of a prestellar core}

%\correspondingauthor{Jorma Harju}
%\email{jorma.harju@helsinki.fi, harju@mpe.mpg.de}

\author[0000-0002-1189-9790]{Jorma Harju}
\affiliation{Max-Planck-Institut f\"ur extraterrestrische Physik,
  Gie{\ss}enbachstra{\ss}e 1, 85748 Garching, Germany}
\affiliation{Department of Physics, P.O. BOX 64, 00014 University
  of Helsinki, Finland}

\author[0000-0002-3972-1978]{Jaime E. Pineda}
\affiliation{Max-Planck-Institut f\"ur extraterrestrische Physik,
  Gie{\ss}enbachstra{\ss}e 1, 85748 Garching, Germany}

\author[0000-0003-1684-3355]{Anton I. Vasyunin}
\affiliation{Max-Planck-Institut f\"ur extraterrestrische Physik,
  Gie{\ss}enbachstra{\ss}e 1, 85748 Garching, Germany}
\affiliation{Ural Federal University, 620002, 19 Mira street, Yekaterinburg, 
Russia}
\affiliation{Ventspils University of Applied Sciences, In{\v{z}}enieru 101, Ventspils 3601, Latvia}

\author[0000-0003-1481-7911]{Paola Caselli}
\affiliation{Max-Planck-Institut f\"ur extraterrestrische Physik,
  Gie{\ss}enbachstra{\ss}e 1, 85748 Garching, Germany}

\author{Stella S.R. Offner}
\affiliation{Astronomy Department, University of Texas, Austin, 
TX 78712, USA}

\author{Alyssa A. Goodman}
\affiliation{Harvard-Smithsonian Center for Astrophysics, 60 Garden Street,
    Cambridge MA 02138, USA}

\author[0000-0002-5809-4834]{Mika Juvela}
\affiliation{Department of Physics, P.O. BOX 64, 00014 University
  of Helsinki, Finland}

\author{Olli Sipil{\"a}}
\affiliation{Max-Planck-Institut f\"ur extraterrestrische Physik,
  Gie{\ss}enbachstra{\ss}e 1, 85748 Garching, Germany}

\author[0000-0001-7199-2535]{Alexandre Faure}
\affiliation{Universit{\'e} Grenoble Alpes, IPAG, F-38000 Grenoble,
  France}
\affiliation{CNRS, IPAG, F-38000 Grenoble, France}

\author[0000-0003-1837-3772]{Romane Le Gal}
\affiliation{Harvard-Smithsonian Center for Astrophysics, 60 Garden Street,
    Cambridge MA 02138, USA}
%\affiliation{Depts. of Chemistry and Astronomy, University of
%  Virginia, McCormick Road, Charlottesville, VA 22904, USA}

\author[0000-0003-3488-8442]{Pierre Hily-Blant}
\affiliation{Universit{\'e} Grenoble Alpes, IPAG, F-38000 Grenoble,
  France}
\affiliation{CNRS, IPAG, F-38000 Grenoble, France}

\author{Jo{\~a}o Alves}
\affiliation{University of Vienna, T{\"u}rkenschanzstra{\ss}e 17, A-1880 Vienna,
  Austria}

\author[0000-0002-9953-8593]{Luca Bizzocchi}
\affiliation{Max-Planck-Institut f\"ur extraterrestrische Physik,
  Gie{\ss}enbachstra{\ss}e 1, 85748 Garching, Germany}

\author[0000-0001-6879-9822]{Andreas Burkert}
\affiliation{Universit{\"a}ts-Sternwarte,
  Ludwig-Maximilians-Universit{\"a}t M{\"u}nchen, Scheinerstra{\ss}e
  1, 81679 M{\"u}nchen, Germany}
\affiliation{Max-Planck-Institut f\"ur extraterrestrische Physik,
  Gie{\ss}enbachstra{\ss}e 1, 85748 Garching, Germany}

\author{Hope Chen}
\affiliation{Harvard-Smithsonian Center for Astrophysics, 60 Garden Street,
    Cambridge MA 02138, USA}

\author[0000-0001-7594-8128]{Rachel K. Friesen}
\affiliation{Dunlap Institute for Astronomy and Astrophysics, University of
    Toronto, 50 St. George Street, Toronto M5S 3H4, Ontario, Canada}

\author{Rolf G{\"u}sten}
\affiliation{Max-Planck-Institut f\"ur Radioastronomie,
  Auf dem H\"ugel 69, 53121 Bonn, Germany}

\author[0000-0002-2885-1806]{Philip C. Myers}
\affiliation{Harvard-Smithsonian Center for Astrophysics, 60 Garden Street,
    Cambridge MA 02138, USA}

\author[0000-0001-6004-875X]{Anna Punanova}
\affiliation{Max-Planck-Institut f\"ur extraterrestrische Physik,
  Gie{\ss}enbachstra{\ss}e 1, 85748 Garching, Germany}
\affiliation{Ural Federal University, 620002, 19 Mira street, Yekaterinburg, 
Russia}

\author{Claire Rist}
\affiliation{Universit{\'e} Grenoble Alpes, IPAG, F-38000 Grenoble,
  France}
\affiliation{CNRS, IPAG, F-38000 Grenoble, France}

\author{Erik Rosolowsky}
\affiliation{ Department of Physics, 4-181 CCIS, University of Alberta,
    Edmonton, AB T6G 2E1, Canada}

\author{Stephan Schlemmer}
\affiliation{I. Physikalisches Institut, Universit{\"a}t zu K{\"o}ln, 
Z{\"u}lpicher Stra{\ss}e 77, 50937 K{\"o}ln, Germany}

\author{Yancy Shirley}
\affiliation{Steward Observatory, University of Arizona, 933 North Cherry
    Avenue, Tucson, AZ 85721, USA}

\author{Silvia Spezzano}
\affiliation{Max-Planck-Institut f\"ur extraterrestrische Physik,
  Gie{\ss}enbachstra{\ss}e 1, 85748 Garching, Germany}

\author{Charlotte Vastel}
\affiliation{Universit{\'e} de Toulouse, UPS-OMP, IRAP, Toulouse, France}
\affiliation{CNRS, IRAP, 9 Avenue du Colonel Roche, BP 44346, F-31028
  Toulouse Cedex 4, France}

\author[0000-0003-2355-4543]{Laurent Wiesenfeld}
\affiliation{Laboratoire Aim\'e-Cotton, CNRS and Universit\'e Paris-Saclay, 91405 Orsay, France}

\begin{abstract}

  We present ALMA maps of the starless molecular cloud core
  Ophiuchus/H-MM1 in the lines of deuterated ammonia (ortho-$\dammo$),
  methanol ($\meth$), and sulphur monoxide (SO). The dense core
  is seen in $\dammo$ emission, whereas the $\meth$ and SO
  distributions form a halo surrounding the core.  Because methanol is
  formed on grain surfaces, its emission highlights regions where
  desorption from grains is particularly efficient. Methanol and
  sulphur monoxide are most abundant in a narrow zone that follows
  the eastern side of the core. This side is sheltered from
    the stronger external radiation field coming from the west. We
    show that photodissociation on the illuminated side can give rise
    to an asymmetric methanol distribution, but that the stark
    contrast observed in H-MM1 is hard to explain without assuming
    enhanced desorption on the shaded side.  The region of the
  brightest emission has a wavy structure that rolls up at one
  end. This is the signature of Kelvin-Helmholtz instability occurring
  in sheared flows.  We suggest that in this zone, methanol and
  sulphur are released as a result of grain-grain collisions induced
  by shear vorticity.

\end{abstract}


\nopagebreak

\section{Introduction} 

\label{sec:intro}

Gaseous methanol ($\meth$) has been found to be present in the
  outer parts of cold starless cores, with abundances of the order of
  $10^{-9}$ relative to $\htwo$ (\citealt{2014A&A...569A..27B};
\citealt{2014ApJ...795L...2V}; \citealt{2016A&A...592L..11S};
\citealt{2016ApJ...830L...6J}; \citealt{2018ApJ...855..112P}). 
  The derived abundances exceed the predictions from pure gas-phase
chemical models by orders of magnitude (\citealt{2006FaDi..133...51G};
\citealt{2016A&A...587A.130B}).  Methanol is believed to form almost
exclusively on the surfaces of dust grains via hydrogenation of frozen
carbon monoxide (CO; \citealt{2002ApJ...571L.173W};
\citealt{2006FaDi..133..177G}), and it is a common constituent of
interstellar ices. To be detectable in the gas phase, methanol must be
released from grains as a result of heating or some non-thermal
mechanism.  In cold, starless cores, non-thermal processes, such
  as impulsive heating by cosmic rays collisions
  (\citealt{1985A&A...144..147L}, \citealt{1993MNRAS.261...83H},
  \citealt{2007MNRAS.382..733R}, \citealt{2018ApJS..239....6K}) and
  desorption caused by exothermic surface reactions
  (\citealt{2006FaDi..133...51G}; \citealt{2007A&A...467.1103G};
  \citealt{2013ApJ...769...34V}; \citealt{2015MNRAS.449L..16B};
  \citealt{2016ApJ...830L...6J}; \citealt{2017ApJ...842...33V})
  provide plausible explanations for the observed fractional
  abundances and distributions of methanol.

It should be noted, however, that core boundaries, where methanol
  is usually found, are also subject to dynamical effects, such as
accretion, velocity shears, and turbulence. Several cores show a sharp
transition from supersonic to subsonic turbulence in a thin layer
surrounding the core (\citealt{1998ApJ...504..223G};
\citealt{2010ApJ...712L.116P}; \citealt{2017ApJ...843...63F};
\citealt{2019ApJ...872..207A}). In this region, the scaling relation
between the velocity dispersion, $\sigma_v$, and the scale length $l$,
$\sigma_v \propto l^a$, seems to break \citep{1998ApJ...504..223G},
suggesting that part of the turbulent energy of the surrounding gas is
dissipated, while at the same time the exterior turbulence compresses
the core. Also gravitational accretion from the surrounding cloud and
collisions between cores can lead to conversion of kinetic energy into
heat. These effects can contribute to the evaporation of the ice
coatings of dust grains.

Photodesorption and photodissociation caused by the external
  radiation field can affect the chemical composition in the outer
  parts of dense cores \citep{2009ApJ...690.1497H}. The effect is
  particularly strong in the vicinity of young massive stars (e.g.,
  \citealt{2018ApJ...861...87G}), and when the core lies near the edge
  of a cloud \citep{2016A&A...592L..11S}. Radiation is also the main
  source of heat at the core boundaries. Finally, cosmic rays, which
  both heat and ionize gas, may also drive chemical changes (e.g.,
  \citealt{2012ApJ...759L..37C}; \citealt{2013A&A...559A..53M};
  \citealt{2016ApJ...819...13C}). While the cosmic ray flux is often
  assumed to be uniform throughout star-forming regions, cosmic rays
  can be accelerated locally by protostellar jets and accretion
  shocks, causing the cosmic ray flux to vary spatially on core and
  cloud scales (\citealt{2016A&A...590A...8P}).  This, in turn,
  produces chemical gradients in dense gas
  (\citealt{2014ApJ...794..123C}; \citealt{2019ApJ...878..105G}).

Here we present maps of a nearby prestellar core in the spectral lines
of methanol, deuterated ammonia ($\dammo$), and sulphur monoxide (SO),
obtained using the Atacama Large (Sub)millimeter Array (ALMA). We
discuss the origin of gas-phase methanol based on the observed
molecular distributions and the physical conditions of the emission
regions. In addition to cosmic-ray collisions and reactive
  desorption, we identify grain-grain collisions induced by gas
  velocity fluctuations as a possible mechanism releasing methanol from
  grains. In this connection we consider grain acceleration in
  three-dimensional turbulence and in shear vorticity. The influence
  of the external radiation field on the asymmetries of the
  $\meth$ and SO distributions is also discussed. 

\begin{table*}[tb]
\caption{Spectral lines observed toward Ophiuchus/H-MM1 with ALMA.}
\label{table:obslines}
{\begin{center}
\begin{tabular}{llcrcrc}
 & & & & & & \\
molecule & transition & frequency & $E_{\rm u}$ & $A_{\rm ul}$ & $\Delta\,v$ & rms$^a$ \\
         &            &   (MHz)    & (K)       &  ($10^{-5}\,\pers$) & ($\ms$) & (K) \\ \hline
o-$\dammo$ & $1_{11}-1_{01}$ & 85926.27 & 20.7 & 0.78 & 107  & 0.042 \\
$\meth$ & $2_{-1}-1_{-1}$ E & 96739.36 & 12.5 & 0.26 &  95  & 0.049  \\
        & $2_{0}-1_{0}$  A & 96741.38 &  7.0 & 0.34  &      &      \\
        & $2_{0}-1_{0}$  E & 96744.55 & 20.1 & 0.34  &      &      \\
        & $2_{1}-1_{1}$  E & 96755.51 & 26.9 & 0.26  &      &       \\
SO      & $2_3-1_2$        & 99299.87  & 9.2   & 1.13 & 1474 & 0.011 \\ \hline
\multicolumn{7}{l}{$^a$ beamsize $4\arcsec$} 
\end{tabular}

\end{center}
}   
\end{table*}

\section{Observations}

\label{sec:observations}

The target of the present observations is the nearby prestellar core
Ophiuchus/H-MM1 located on the eastern side of the L1688 cloud
(\citealt{2004ApJ...611L..45J}; \citealt{2011A&A...526A..31P};
\citealt{2017A&A...600A..61H}). The dimensions of the kidney-shaped
dense core are approximately $1\arcmin\times 2\arcmin$.  The
target is prominent in the $850\,\mu$m dust continuum maps of
Ophiuchus observed with SCUBA-2 \citep{2015MNRAS.450.1094P}, and
in the $\ammo$ map of L1688 from the Green Bank Ammonia Survey
\citep{2017ApJ...843...63F}.

The present data were taken during the ALMA cycle 4 (project
2016.1.00035.S).  Here we discuss the $J_k=2_k - 1_k$ rotational lines
of $\meth$ at 96.7 GHz and the $J_{K_a,K_c}=1_{11}^{\rm s}-1_{01}^{\rm
  a}$ rotation-inversion line of ortho-$\dammo$ at 85.9 GHz, which
were observed simultaneously with the 3 millimeter continuum in the
ALMA Band 3. The 'continuum' spectral window included the
$J_N=3_2-2_1$ rotation line of SO at 99.3\,GHz. This line is
unresolved because the channel width in this spectral cube is $\sim
4.9$\,MHz ($\sim 1.5\,\kms$, see Table~\ref{table:obslines}). 

An area of $95\arcsec\times135\arcsec$ covering the brightest
  $850\,\mum$ emission was imaged using the ALMA 12\,m array (40
antennas) in one of its most compact configurations, and the ALMA
Compact Array (ACA) with 10 7\,m antennas.  The total power (TP)
antennas were not used. With the 12\,m array, the mapping was carried
out by a five-point mosaic, whereas with the 7\,m array, a single
point was measured. The data were calibrated and imaged using the CASA
version 4.7.2. The synthesized beam sizes of naturally weighted
  images ranged from $3\farcs8\times2\farcs5$ (at 85.9\,GHz) to
  $3\farcs5\times2\farcs3$ (at 99.3\,GHz). The largest recoverable
  scale of the images is approximately $60\arcsec$. In what follows,
  we use image cubes that were created using a common, circular
  $4\arcsec$ restoring beam for all lines. Based on recent
    distance determinations for L1688 using the VLBA and {\sl Gaia}
    (\citealt{2017ApJ...834..141O}; \citealt{2018ApJ...869L..33O}) we
      adopt a distance of 140\,pc for the core. The angular resolution
      $4\arcsec$ corresponds to 560\,au at this distance. The observed
    transitions are listed in Table~\ref{table:obslines}. The rms
    noise values in this Table are given for the image cubes with
    $4\arcsec$ resolution. For the 3\,mm continuum map, the rms noise
    is approximately 0.06\,mJy/beam.

\section{Integrated intensity maps}

\label{sec:maps}

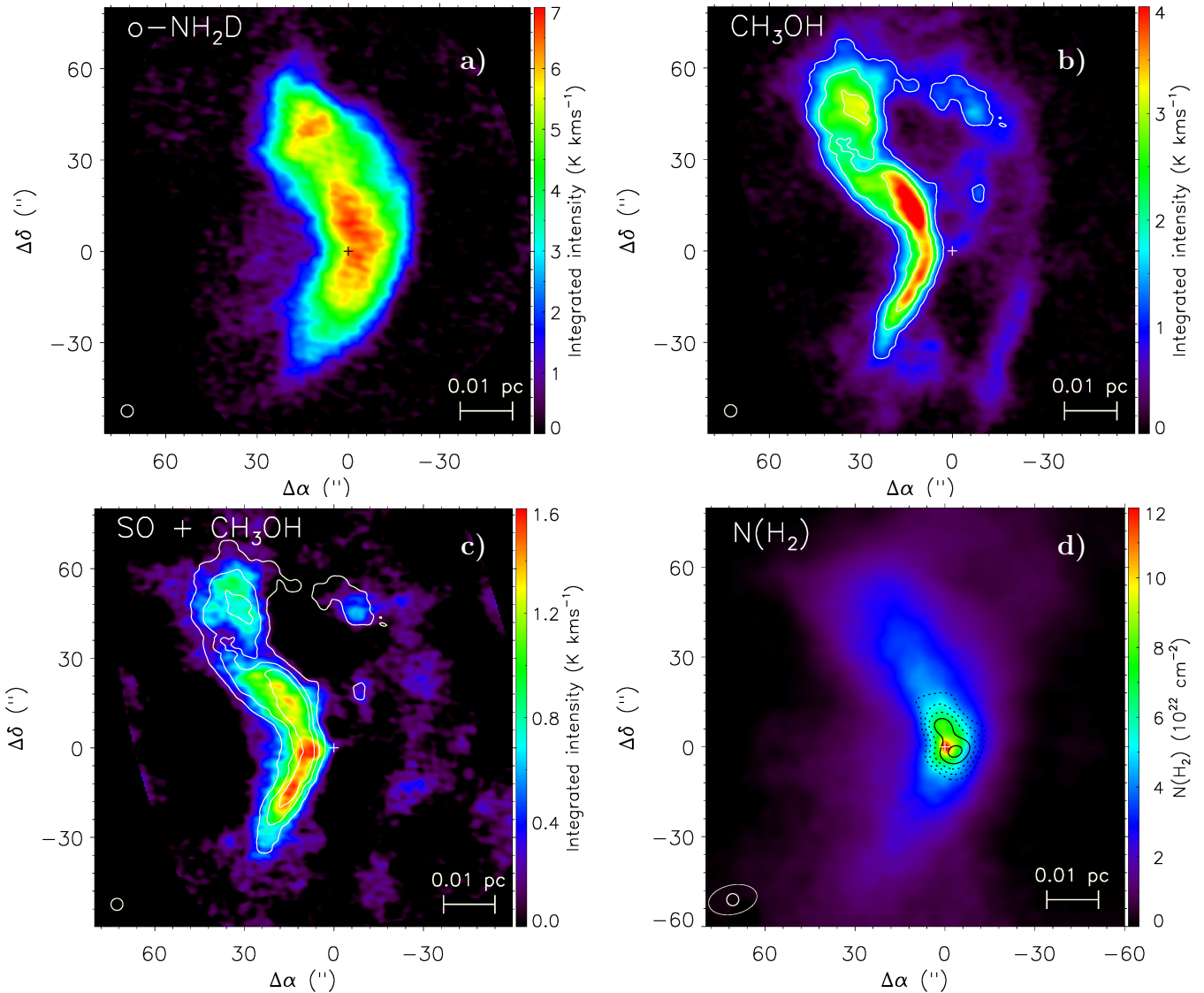
\begin{figure*}[]
\unitlength=1mm
\begin{picture}(160,145)(0,0)
\put(-2,72){
\begin{picture}(0,0) 
\includegraphics[width=9.0cm,angle=0]{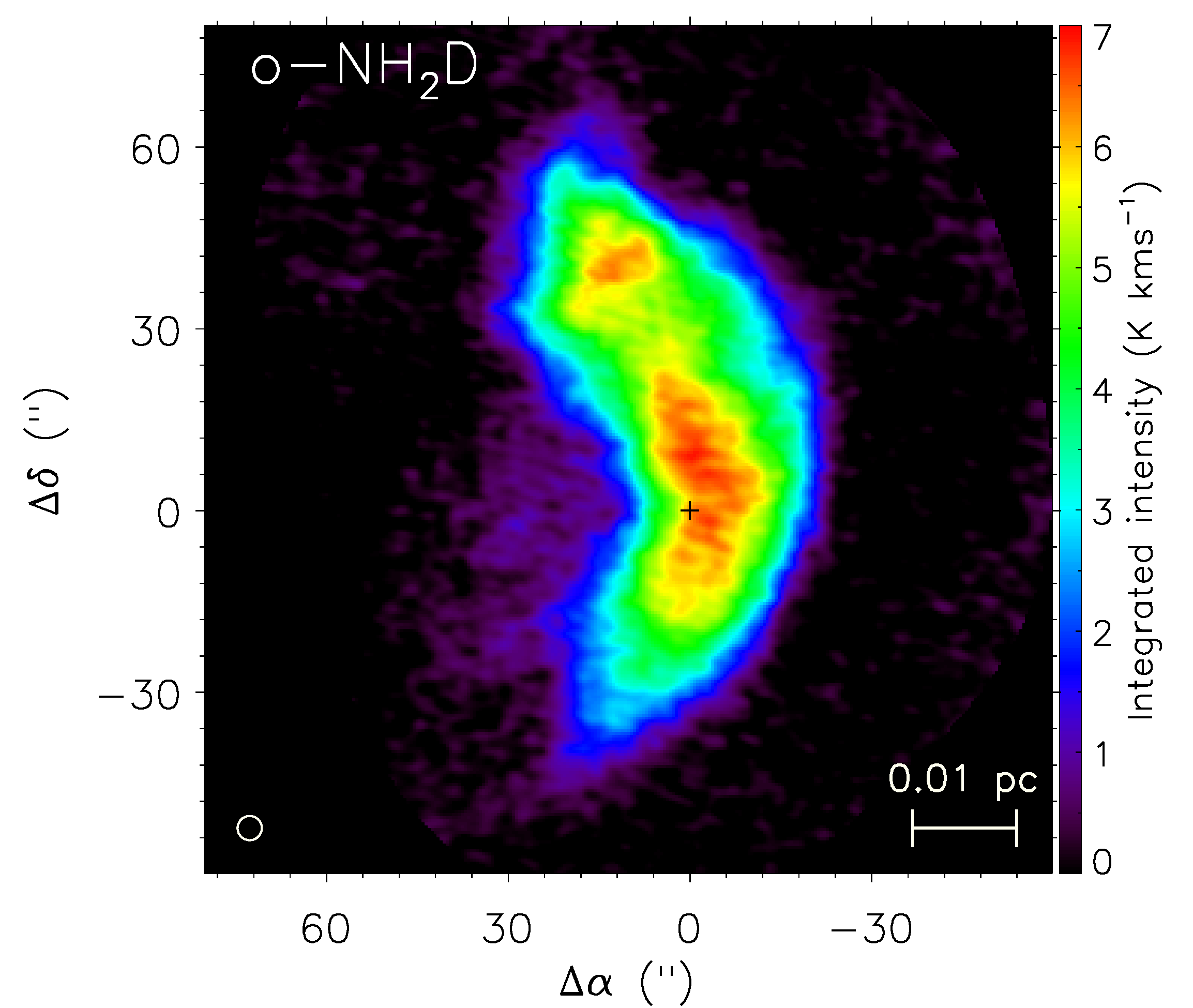} 
\end{picture}}
\put(90,72){\begin{picture}(0,0) 
\includegraphics[width=9.0cm,angle=0]{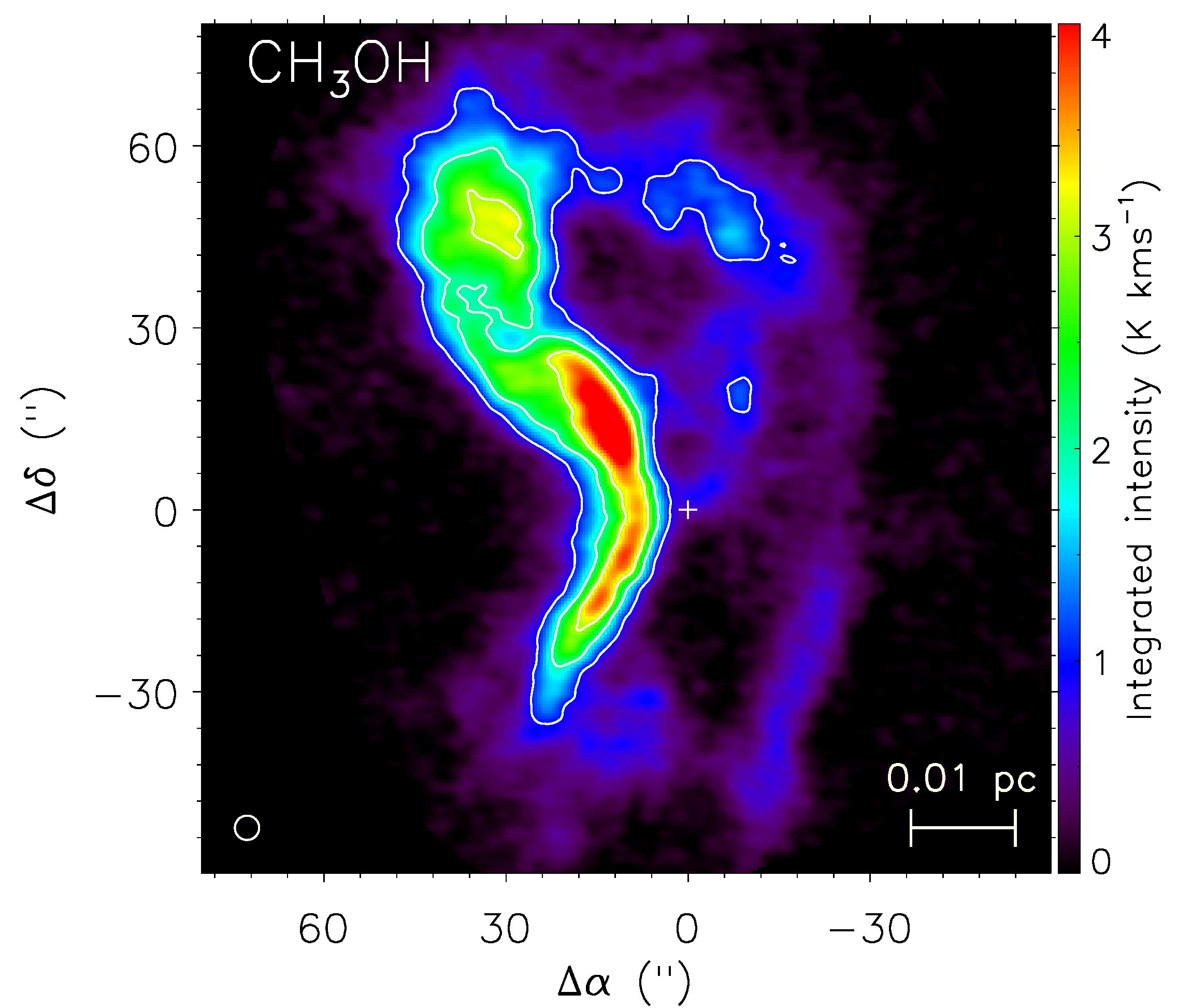}
 \end{picture}}
\put(-2,-2){\begin{picture}(0,0) 
\includegraphics[width=9.0cm,angle=0]{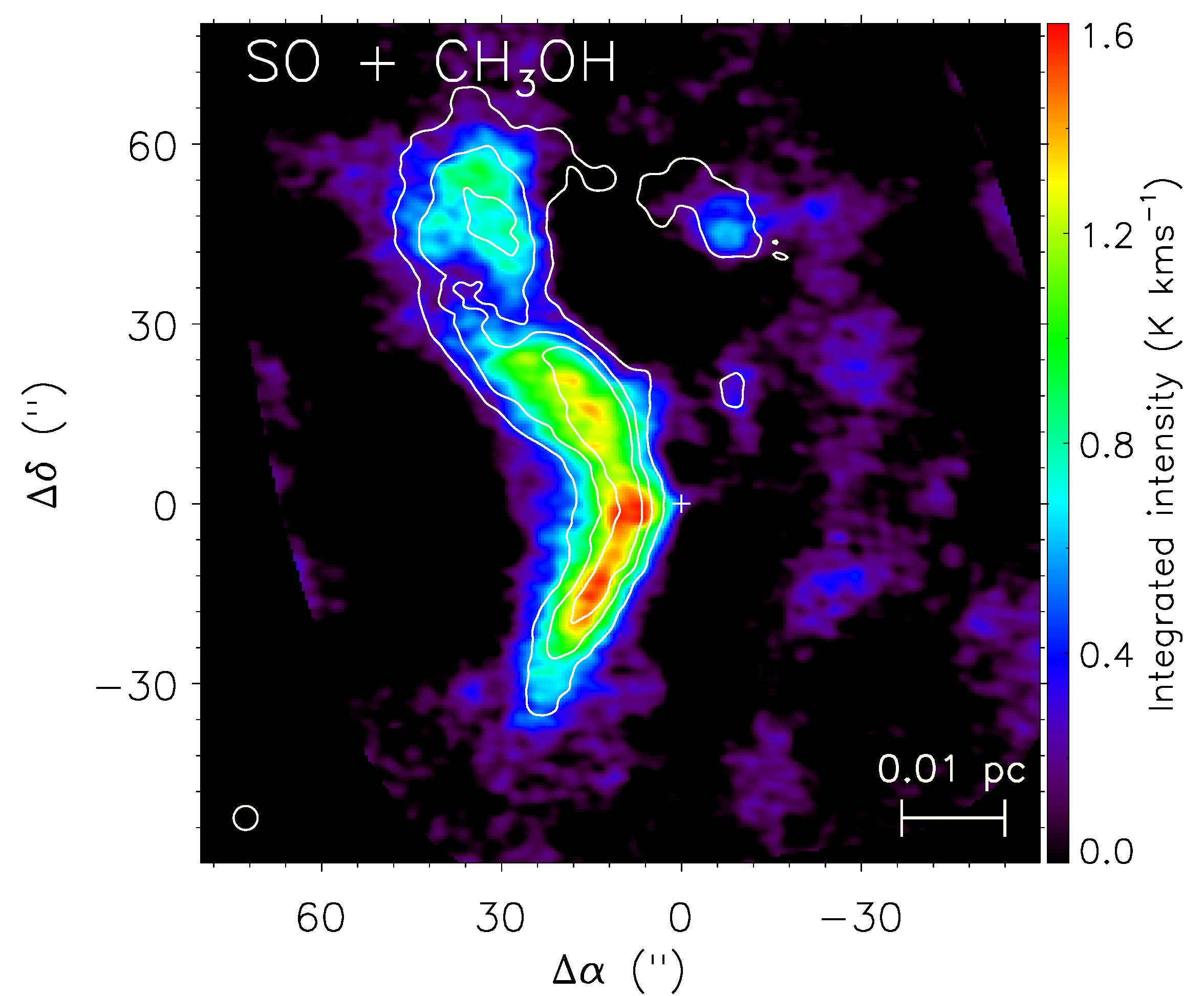}
 \end{picture}}
\put(90,-2){\begin{picture}(0,0) 
\includegraphics[width=9.0cm,angle=0]{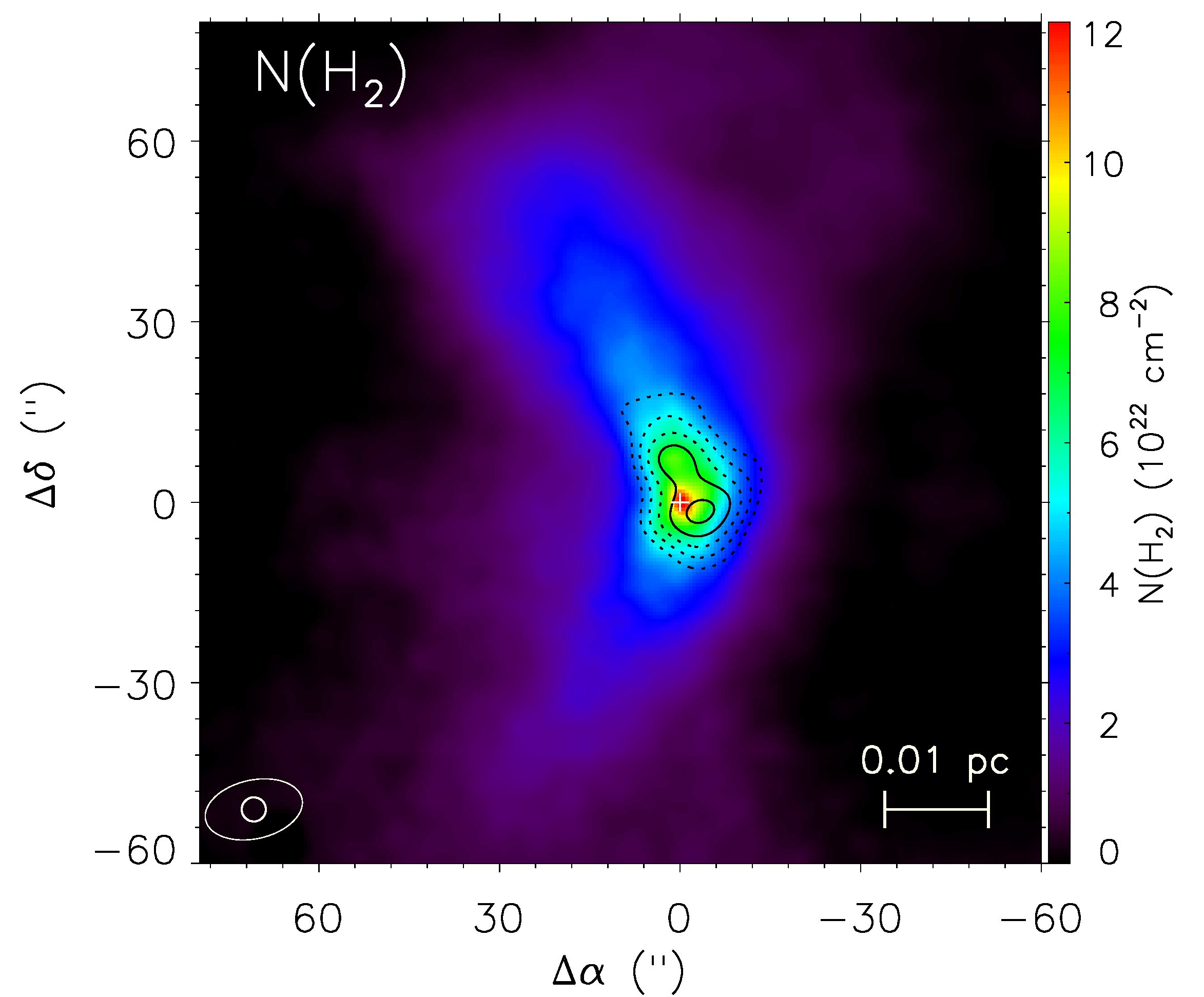}
 \end{picture}}
\put(70,138){\makebox(0,0){\large \bf \color{white} a)}}
\put(160,138){\makebox(0,0){\large \bf \color{white} b)}}
\put(70,65){\makebox(0,0){\large \bf \color{white} c)}}
\put(160,65){\makebox(0,0){\large \bf \color{white} d)}}
\end{picture}
\caption{Molecular line maps and the $\htwo$ column density map of the
  Ophiuchus/H-MM1 core. The panels (a), (b) and (c)
  show the integrated intensity maps of the
  ortho-$\dammo(1_{11}-1_{01})$, $\meth(2_k-1_k)$ and SO$(2_3-1_2)$
  lines measured by ALMA. Methanol contours are superposed on the
    SO map in panel (c).  The contour levels are 1, 2, and
    $3\,\Kkms$. The synthetic beam size of the ALMA images
  ($4\arcsec$) is indicated in the bottom left of each panel.  The
  $N(\htwo)$ map of panel (d) is derived from 8\,$\mum$
  extinction, based on observations by the Spitzer Space
  Telescope. The contour lines show the 3\,mm continuum emission
  measured by the ACA. The levels go from 2 to 3 mJy\,beam$^{-1}$ in
  steps of 0.25\,mJy\,beam$^{-1}$. The ellipse in the bottom left
  represents the synthetic beam of the ACA map. The plus sign
  indicates the position of the column density maximum with the
  coordinates RA 16:27:58.65, Dec. -24:33:41.2 (J2000).}
\label{figure:line_maps}
\end{figure*}

The integrated intensity maps of the ortho-$\dammo$, $\meth$ and SO
lines at 85.9, 96.7 and 99.3\,GHz, respectively, are shown in
Figures~\ref{figure:line_maps}a, b and c. The methanol map is
  also shown in panel c as contours superposed on the SO map. The
fourth map, shown in panel d of this figure, is the $\htwo$ column
density map derived from 8\,$\mum$ extinction. For this we have used
the 8\,$\mum$ surface brightness map measured by the InfraRed Array
Camera (IRAC) of the Spitzer Space Telescope, smoothed to a $4\arcsec$
resolution (the original resolution is $\sim 2\arcsec$). The method
used for deriving the $N(\htwo)$ map is described in
Appendix~\ref{sec:column}. The 3\,mm continuum emission is weak with a
  peak intensity of $\sim0.3$\,mJy/beam, that is, 5 times the rms
  noise of the map. The continuum sensitivity of the present ALMA
  observations was not sufficient for detecting extended emission
  around this peak. The 3\,mm continuum map using only the ACA data
is shown in Figure~\ref{figure:line_maps}d, as a contour plot
superposed on the $N(\htwo)$ image.

The $\dammo$ map resembles the $\htwo$ column density map shown in
Figure~\ref{figure:line_maps}d, and the far-infrared dust emission
maps of the core, e.g., the 850\,$\mu{\rm m}$ SCUBA-2 map shown in
Figure~\ref{figure:taumidfar}. The $\meth$ and SO distributions
closely resemble each other (see Figure~\ref{figure:line_maps}c),
and they are almost complementary to the
$\dammo$ distribution; methanol and sulphur monoxide follow the edges
of the elongated core but are stronger on the left, concave side of
the core (the eastern side in the sky) than elsewhere.

\section{Column densities and velocity distributions}

\label{sec:line_analysis}

Four lines of the $(2_k-1_k)$ group of $\meth$ were included in the
spectral window around 96.7 GHz. Three of them belong to the $E$
symmetry species and one to the $A$ species. The rotational
temperatures, $T_{\rm rot}$, and the column densities, $N$, of
$E$-type methanol were derived adopting the procedure described by
\cite{2000ApJS..128..213N}, where no assumption about the optical
thickness of the lines is made. The calculation was limited to
  the positions where at least two of the $E-\meth$ lines are detected
  with $3\sigma$ significance. In practice this means positions where
  the integrated intensity of the $2_0-1_0$ line is greater than
  $\sim 0.045\,{\rm K}\,\kms$. The region where this condition holds
  is contained within the $1\,{\rm K}\,\kms$ contour of the total
  integrated intensity map shown in Figures~\ref{figure:line_maps}b
  and c.  The weighted average of the rotational temperature and its
  standard deviation are ${\bar T}_{\rm rot} = 10.3\pm0.7$\,K. The
  peak value of the $E$-methanol column density is
  $N(E-\meth)=1.1\pm0.1\times10^{14}\,\persqcm$.  In the region with
  the brightest $\meth$ emission, bordering the eastern side of the
  core, $N(E-\meth) \geq 5\times10^{13}$ cm$^{-2}$. On the western
  side, the column density could only be determined toward irregularly
  dispersed positions. The values in these positions are typically
  less than $\sim 5\times10^{12}\,\persqcm$, except for a couple of positions 
  reaching $\sim 2\times10^{13}\,\persqcm$. The $A$-methanol column
density, $N(A-\meth)$, was estimated using the single $A$-line in the
spectrum, assuming that the rotational temperatures of the $A$ and $E$
symmetry species are the same. The weighted average of the $A/E$
  ratio and its standard deviation are $1.3\pm0.2$ (the
high-temperature statistical value is 1.0). This indicates that
$N({\meth}) \sim 2\times N(E-\meth)$.
 
The data set contains only one, spectrally unresolved SO line,
$J_N=3_2-2_1$, at 99.3 GHz. Therefore, we only can derive the lower
limit of the SO column density assuming optically thin emission.  In
this approximation, the upper state column density is directly
proportional to the integrated brightness temperature, $\int\, T_{\rm
  B}\, dv$, of the line. A rotation temperature, assumed to be the
same for all rotational transitions, needs to be adopted for the
calculation of the partition function. Assuming $T_{\rm rot}=10$\,K, we
get $N({\rm SO})=1.2\times10^{13}\,{\rm cm}^{-2} \times \int\, T_{\rm
  B}\, dv$, when the integral is in K\,$\kms$. For $T_{\rm rot}=5$\,K,
the numerical factor before the integral is $1.7\times10^{13}\,{\rm
  cm}^{-2}$.

\begin{figure}[htb]
\unitlength=1mm
\begin{picture}(80,63)(0,0)
\put(0,-2){
\begin{picture}(0,0) 
\includegraphics[width=8.0cm,angle=0]{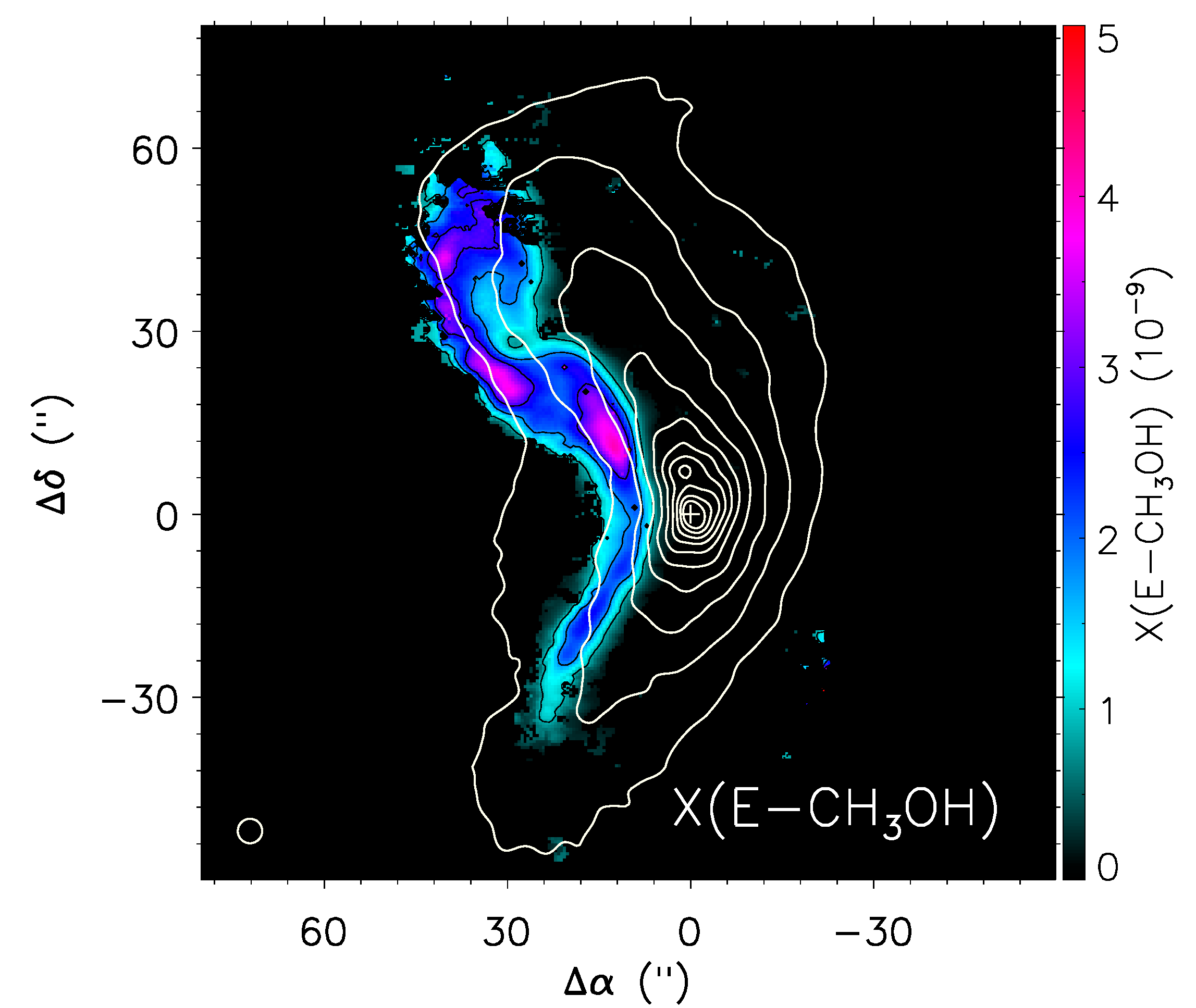} 
\end{picture}}
\end{picture}
\caption{Fractional $E-\meth$ abundance distribution. The contours represent the $\htwo$ column density. They go from $1\times10^{22}$\,cm$^{-2}$ to 
$10\times10^{22}$\,cm$^{-2}$ in steps of $1\times10^{22}$\,cm$^{-2}$.}
\label{figure:xmeth}
\end{figure}

The fractional $\meth$ and SO abundance distributions were determined
by dividing the column density maps by the $N(\htwo)$ map shown in
Figure~\ref{figure:line_maps}d. The fractional $\meth$ abundance map
is shown in Figure~\ref{figure:xmeth}. 
The highest methanol abundances, $X(\meth)\sim 8\times10^{-9}$
(assuming equal abundances for the $E$ and $A$ types) are found at the
eastern border of the core. The SO abundances are highest south of the
methanol peak, with $X({\rm SO}) \sim 0.7-1.0\times 10^{-9}$, depending
on the assumed $T_{\rm rot}$ (10 or 5 K).

In this paper, we only use the kinematic information from the $\dammo$
lines; the $\dammo$ column densities and fractional abundances will be
discussed elsewhere. The ortho-$\dammo$ spectral cube was analyzed by
performing multicomponent Gaussian fits to the hyperfine structure of
the $1_{11}-1_{01}$ line which consists of 5 separate groups of
  hyperfine components; four satellites located symmerically on both
  sides of the strongest main group.  The model used for the
hyperfine structure takes into account the splittings owing to the
electric quadrupole moments of both N and D nuclei
\citep{2016A&A...586L...4D}.  The fits were made to positions where
each of the outer satellites are detected with $3\sigma$ significance, 
meaning that their integrated intensity is greater than 
 $\sim 0.050\,{\rm K}\,\kms$.
These fits give accurate estimates for the line velocity and width.
The velocity distribution of methanol was derived by Gaussian
  fits to the four lines near 96.7\,GHz. In this fitting, the line
  separations were fixed and it was assumed that the velocity
  dispersion is the same for all components. 

\begin{figure*}[htb]
\unitlength=1mm
\begin{picture}(150,120)(0,0)
\put(0,0){
\begin{picture}(0,0) 
\includegraphics[width=18.0cm,angle=0]{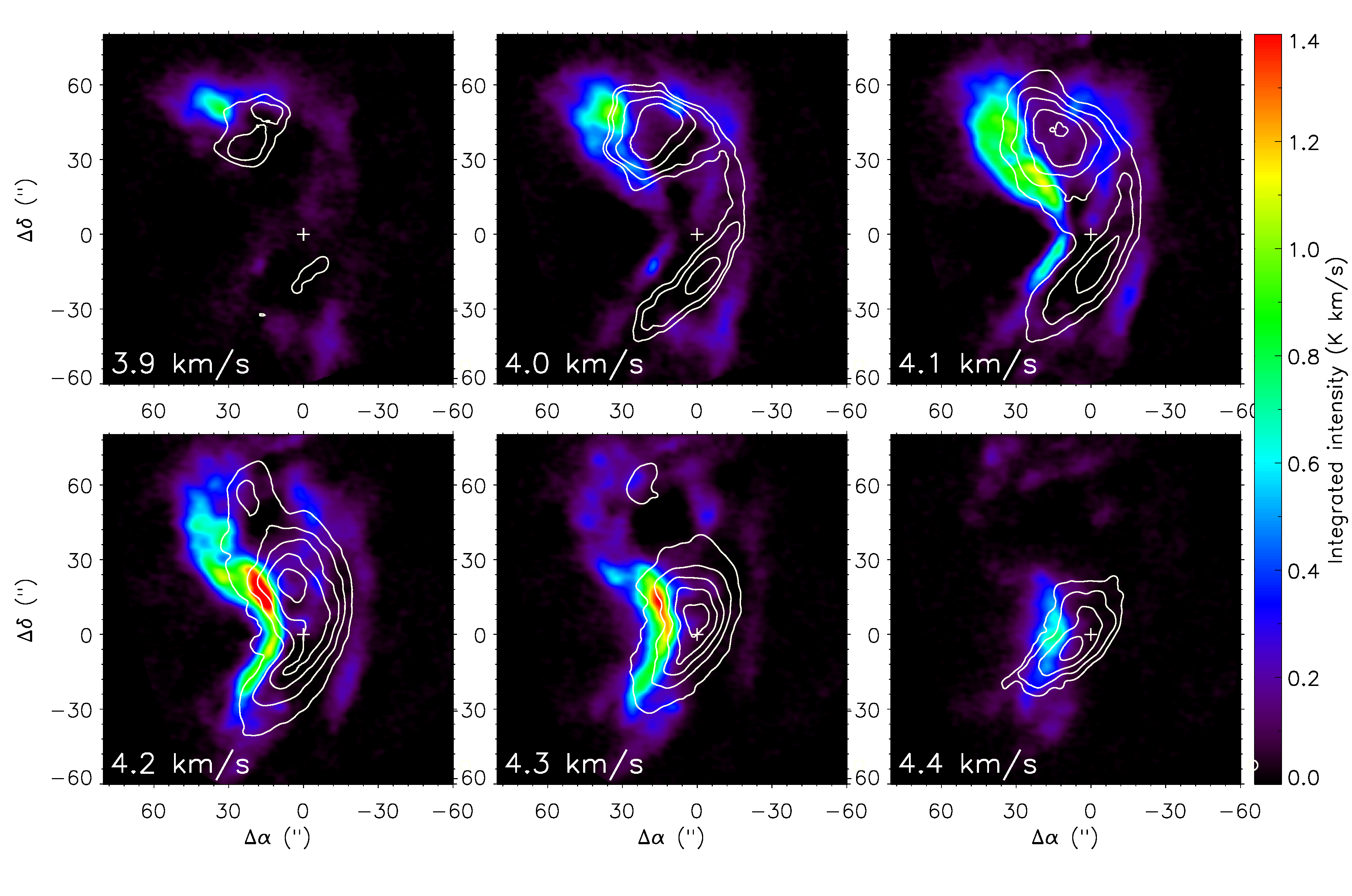} 
\end{picture}}
\end{picture}
\caption{Velocity channel maps of $\meth$ and $\dammo$ (contours) emissions.
  The contour levels of the $\dammo$ maps are: 0.2 and 0.4\,$\Kkms$
    (3.9\,$\kms$); 0.2, 0.4, 0.8 and 1.4\,$\Kkms$ (4.0 and 4.4\,$\kms$); 0.2, 0.4, 0.8, 1.4, 2.0 and 2.6\,$\Kkms$ (4.1-4.3\,$\kms)$.}
\label{figure:channel_maps}
\end{figure*}

\begin{figure*}[]
\unitlength=1mm
\begin{picture}(150,75)(0,0)
\put(-2,-2){\begin{picture}(0,0) 
\includegraphics[width=9.0cm,angle=0]{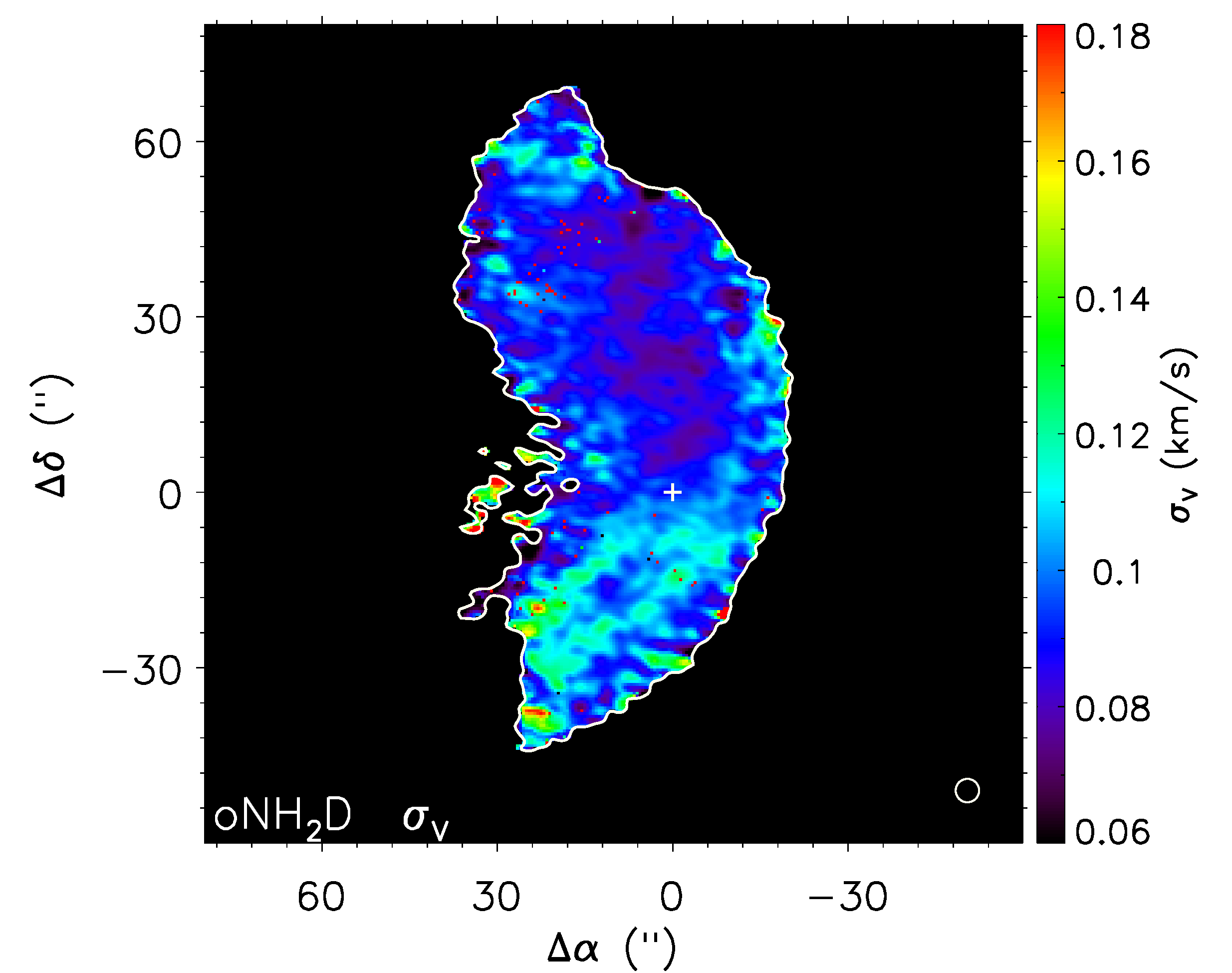}
 \end{picture}}
\put(90,-2){\begin{picture}(0,0) 
\includegraphics[width=9.0cm,angle=0]{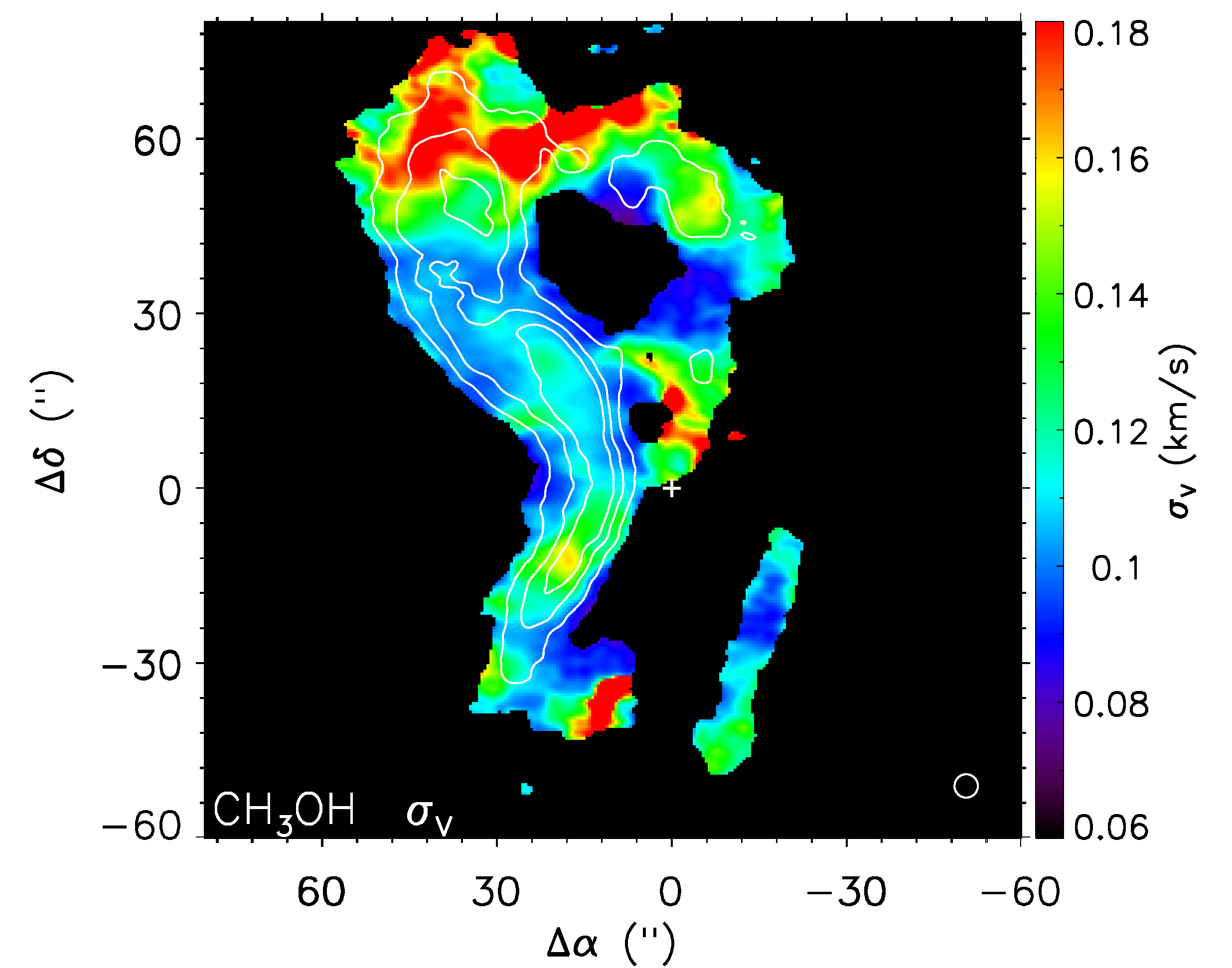}
 \end{picture}}
\put(70,65){\makebox(0,0){\large \bf \color{white} a)}}
\put(160,65){\makebox(0,0){\large \bf \color{white} b)}}
\end{picture}
\caption{Velocity dispersions of $\dammo$ (a) and $\meth$ (b) spectral lines.}
\label{figure:sigma_maps}
\end{figure*}

Multicomponent Gaussian fits to the $\dammo$ and $\meth$ lines
  were used to produce collapsed line profiles, that is, single
  Gaussians with the same integrated intensity, radial velocity, and
  velocity dispersion as the original lines consisting of several
  frequency components. Radial velocity channel maps over the velocity
  range $3.9\,\kms$ to $4.4\,\kms$ derived from the collapsed $\dammo$
  and $\meth$ profiles are shown in Figure~\ref{figure:channel_maps}
  (the full range of detectable emission is $3.7-4.6\,\kms$). The velocity
  dispersion maps are shown in Figure~\ref{figure:sigma_maps}. A
  similar analysis was not possible for the unresolved SO spectra.

\begin{figure}[htb]
\unitlength=1mm
\begin{picture}(80,55)(0,0)
\put(0,0){
\begin{picture}(0,0) 
\includegraphics[width=8.0cm,angle=0]{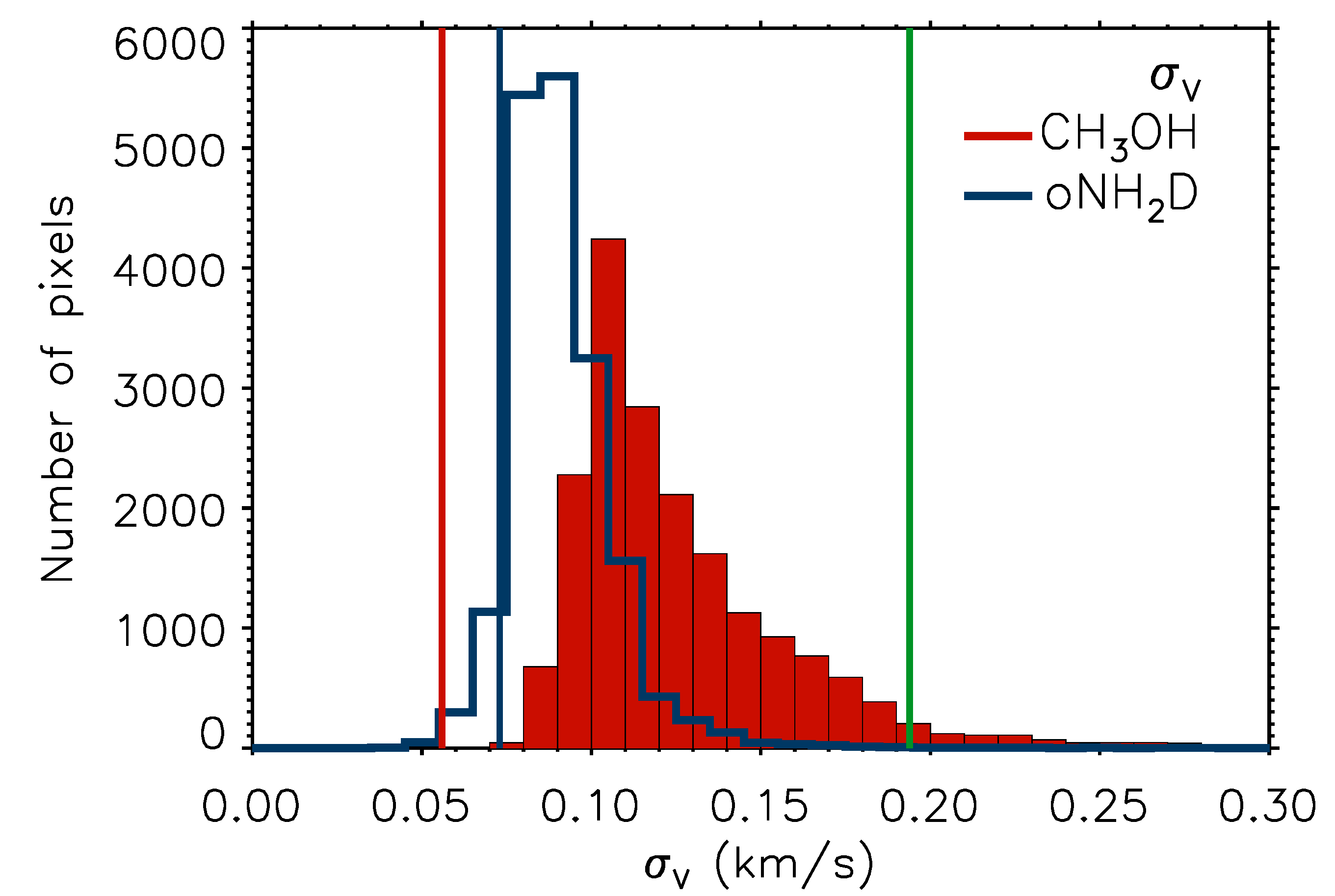} 
\end{picture}}
\end{picture}
\caption{One-dimensional velocity dispersions of the ortho-$\dammo$
  and $\meth$ lines. The blue and red vertical lines
  indicate the thermal velocity dispersions of $\dammo$ and $\meth$ at
  10 K. The green vertical line indicates the sound speed at this
  temperature.}
\label{figure:sigma_histos}
\end{figure}

The channels maps of Figure~\ref{figure:channel_maps} show an
  overall north-south gradient. The velocity change (including all
  channels with detectable emission) is approximately $0.7\,\kms$ over
  a distance of 0.034\,pc, corresponding to $50\arcsec$ in the
  sky. The core motion is reminiscent of an eddy rather than rigid
  body rotation, and seems to support the idea of core formation in
  turbulent converging flows \citep{2011ApJ...729..120G}.  Despite an
  east-west offset, $\meth$ and $\dammo$ follow the same velocity
  pattern, and it is remarkable how the $\dammo$ peak in each channel
  is found some $10\arcsec - 20\arcsec$ west of the $\meth$ peak.
  Methanol lines, originating at the core boundaries, have larger
  line-of-sight velocity dispersions than $\dammo$ lines
  (Figure~\ref{figure:sigma_maps}). Histograms of the one-dimensional
  velocity dispersions of the lines are shown in
  Figure~\ref{figure:sigma_histos}. This figure indicates that the
  line widths of $\dammo$ are nearly thermal, assuming a kinetic
  temperature of 10\,K, whereas $\meth$ lines have a substantial
  non-thermal component. However, also for $\meth$ the velocity
  dispersion is subsonic in most positions.

The relationship between the $\meth$ and $\dammo$ distributions
  in space and velocity is visualized in a video available at
  \url{https://www.youtube.com/watch?v=C814sOPF7c0}. This video
  demonstrates the superimposition of the $\meth$ and $\dammo$ data
  cubes using Glue\footnote{\url{https://glueviz.org}} which is an
  open-source Python library designed for the analysis of
  multi-dimensional related datasets. The close complementarity of the 
two distributions is particularly evident in this video.

\section{Note on the $\dammo$ distribution}

\label{sec:dammo}

 Ammonia and its deuterated isotopologues form both in the gas and
  on grains, and are subject to accretion and desorption just like
  $\meth$ and SO. Why is the distribution of $\dammo$ then so
  different from those of $\meth$ and SO?  One thing that affects this
  difference is that ammonia (along with some other nitrogen
  containing molecules; \citealt{2010A&A...513A..41H}) is less
  susceptible to freeze-out than most C- and O-bearing molecules. This
  observational fact is not fully understood
  \citep{2019MNRAS.487.1269S}.  The persistence of gaseous ammonia is
  evident, for example, from the Green Bank Ammonia Survey maps of
  nearby molecular clouds (\citealt{2017ApJ...843...63F}; see also the
  maps in \citealt{2006A&A...455..577T}). \cite{2017A&A...600A..61H}
  suggested that one of the reasons for this persistence is the
  chemical inertia of N$_2$ molecules on grain surfaces, which makes
  them return to the gas phase rather than form chemical bonds on the
  surface.

  The other reason for the contrasting distributions of $\dammo$ and
  $\meth$ is that deuterium fractionation starts in earnest at high
  densities where CO has largely disappeared from the gas, whereas
  $\meth$ production (through desorption) is probably most active
  further out in a starless core, where CO is not severely depleted
  (see Section~\ref{sec:desorption}).  In the gas phase, deuterated
  ammonia is mainly produced in reactions between $\ammo$ and
  deuterated ions (\citealt{2001ApJ...553..613R};
  \citealt{2015A&A...581A.122S}; \citealt{2018MNRAS.477.4454H}).
  Deuterated ions are enhanced as a consequence of the freeze-out of
  CO, which first leads to a rapid increase of $\hthree$. Deuterium
  fractionation is boosted when the fractional CO abundance decreases
  below that of HD, $\sim 3\times10^{-5}$.  The primary fractionation
  reaction is $\hthree + \hd \rightarrow \htwod + \htwo$. Successive
  reactions with HD produce also $\dtwoh$ and $\dthree$. In cold,
  dense clouds, the deuterated forms of $\hthree$ compete for the
  position of dominant ion and transfer deuterium to other species
  including $\ammo$ (\citealt{2003ApJ...591L..41R};
  \citealt{2004A&A...418.1035W}; \citealt{2019RSPTA.37780401C}).

  Ammonia attains high abundances on grains, where it is formed
  through H atom additions to N atoms. The formation of $\dammo$ and
  other deuterated molecules on grains thus requires free D atoms
  which usually have very low abundances in molecular clouds as almost
  all deuterium is locked in HD.  The abundance of free D atoms
  increases, however, significantly when deuterium chemistry
  flourishes in the gas phase. The principal source of deuterium atoms
  is the dissociative electron recombinations of the $\htwod$,
  $\dtwoh$ and $\dthree$ ions, which benefit from the depletion of
  CO. The production of $\dammo$ is therefore strongly favoured inside
  the CO depletion zone. The predicted radial distributions of the
  fractional abundances for different ammonia isotopologs in a dense
  starless core model resembling H-MM1 are shown in Fig.~14 of
  \cite{2017A&A...600A..61H}. Because the $\htwo$ density distribution
  is centrally peaked, also the $\dammo$ column density is highest at
  the core center. In H-MM1, the detectable $\dammo$ emission is
  confined to the dense region which is seen as a dark patch on the 8 and
  $24\,\mum$ surface brightness maps (see Appendices~\ref{sec:column}
  and \ref{sec:tdust}).

\section{Desorption of methanol}

\label{sec:desorption}

In this Section, we describe mechanisms that can release methanol from
grains and estimate their efficiencies in a cold, dense cloud. This is to
provide background for the next Section where we discuss the origin of
the asymmetric methanol distribution observed in H-MM1 with ALMA.

The binding energy of methanol on a water ice surface is
  approximately 5,500\,K (0.47 eV), and
  its thermal evaporation from interstellar grains is likely to become
  significant at around 100\,K where also $\water$ and $\ammo$
  sublimate \citep{2006A&A...457..927G}. In the interiors of dense
  starless clouds, the otherwise cold dust grains can be impulsively
  heated by various mechanisms, leading to the desorption of methanol
  and other species formed on their surfaces. Moveover, UV photons
  created in cosmic ray collisions with $\htwo$ molecules can
  contribute to the desorption of the mantle material.  Also a
  combination of different mechanisms may be effective.  For example,
  impulsive heating by a cosmic ray can trigger a chain of
  radical-radical reactions and an explosive release of chemical
  energy stored in the grain mantle (\citealt{1982A&A...109L..12D};
  \citealt{1985A&A...144..147L}; \citealt{2015ApJ...805...59I}).  In
  the experiments of \cite{1982A&A...109L..12D}, explosive events in
  grain mantle analogs always occurred at $\sim 27$\,K, probably
  because radicals in their samples became mobile at this
  temperature. However, no runaway reaction was observed without
  preceding UV photolysis, which was needed to produce the
  radicals. The number of incident photons in their experiment was
  20\% of the number of sample molecules, and they estimate that a
  ratio of 10\% could still be sufficient for explosive ejection.

A grain accretes constantly atoms and molecules from the gas,
including CO and the highly mobile H atoms, and this alone could be
thought to result in accumulation of radicals in the icy mantle. The
number of radicals formed through accretion is likely to be small,
however, compared to that of saturated molecules because of efficient
hydrogenation and oxidation of surface species.  The characteristic
fraction of reactive molecules has been estimated to be maximally a
few percent (\citealt{1985A&A...144..147L};
\citealt{2015ApJ...805...59I}). Models and observations suggest that
the grain mantle has a layered structure with water ice in the
bottom, covered by layers of CO ice which forms at higher densities
(\citealt{1982A&A...114..245T}; \citealt{2001ApJ...547..872W};
\citealt{2006A&A...453L..47P}). According to the model results of
\cite{2009A&A...508..275C} the mantle surfaces are dominated by
$\meth$ ice in cold dense cores, while CO and other less saturated
species remain locked in the lower ice layers. The complete
hydrogenation of CO to $\meth$ is favoured in conditions where CO
freezes out ($n(\htwo) \ga 10^5\,\percc$, $T \la 15$\,K), depending on
the H/CO abundance ratio in the gas phase \citep{2009A&A...508..275C}.

\subsection{Cosmic ray-induced desorption}

\label{sec:cr_desorption}

Cosmic ray-induced desorption is supposed to sustain traceable
  gas-phase abundances of O- and C-bearing molecules also in the inner
  parts of cold, dense clouds (e.g., \citealt{1985A&A...144..147L};
  \citealt{1993MNRAS.261...83H}; \citealt{2004A&A...415..203S};
  \citealt{2007MNRAS.382..733R}; \citealt{2009ApJ...690.1497H}). The
  efficiency of cosmic ray-induced sputtering of methanol embedded in
  water ice has been studied experimentally by
  \cite{2019A&A...627A..55D}. In their Fig.~11,
  \cite{2019A&A...627A..55D} present estimates for $\meth$ sputtering
  rates as functions of the cosmic ray ionization rate for two
  methanol fractions in the ice, 0.55 and 0.056 with respect to
  $\water$. The lower of these fractions is characteristic of
  interstellar ices \citep{2015ARA&A..53..541B}. Assuming that the
  cosmic ray ionization rate of $\htwo$ is
  $\zeta^{\htwo} \sim 6\times10^{-17}\,\pers$ at the core boundary,
  one obtains from Fig.~11 of \cite{2019A&A...627A..55D} that the flux of
  sputtered $\meth$ molecules is
  $F^{\rm sp}_{\meth} \sim 0.14\,\persqcm\,\pers$ (here we have
  multiplied the value indicated in the figure by two to account for
  the exiting cosmic ray, as suggested in the caption). The
  adopted $\zeta^{\htwo}$ is consistent with the models of
  \cite{2009A&A...501..619P}, in view of the fact that the $\htwo$
  column density of ambient cloud around H-MM1 is
  $\sim 10^{22}\,\persqcm$. 

  In order to estimate the desorption rate per H atom or per cubic
  centimeter in the interstellar medium, we adopt the MRN grain size
  distribution \citep{1977ApJ...217..425M}. This is a power-law
  distribution of spherical grains with
  $\frac{dn_{\rm g}}{da} \propto a^{-3.5}$ between
  $a_{\rm min}=0.005\,\mum$ (50\,\AA) and $a_{\rm max}=0.25\,\mum$,
  where $a$ is the grain radius and $n_{\rm g}$ is the number density
  of grains. Assuming that the dust-to-gas mass ratio is 0.01 and that
  the average grain material density is $3.0\,{\rm g}\,\percc$, the
  cross-sectional area of dust grains is
  $\sigma_{\rm H}=1.64\times 10^{-21}\,\sqcm$ per H
  atom. %The total surface area of the grains is 4 times larger.
  The cosmic ray-induced sputtering rate of $\meth$ from water ice is
  then $\sim2.3\times10^{-22}$ molecules\,$\pers$ per H atom. The rate
  per cubic centimeter and second can be obtained by multiplying this
  number by the total hydrogen density $n_{\rm H}$, which is
  $\sim 2\times n(\htwo)$ in molecular clouds. The corresponding
  desorption rate per methanol molecule is
  $\sim5.7\times10^{-17}$\,mol$^{-1}\,\pers$, assuming that the
  fractional methanol abundance on grains relative to H atoms is
  $n(\meth,{\rm s})/n_{\rm H} \sim 4\times 10^{-6}$.  This abundance
  is derived using the composition of the ice mantles in quiescent
  clouds and cores listed in Table~2 of \cite{2015ARA&A..53..541B},
  assuming that the mantle constitutes 15\% of the grain mass.  The
  adopted $\meth$ abundance agrees reasonably well with gas-grain
  chemistry models (e.g., \citealt{2014ApJ...791....1T};
  \citealt{2017ApJ...842...33V}). 
  \cite{2019A&A...627A..55D} estimate that the flux of sputtered
$\meth$ from CO ice is 40 times higher than that from water ice, which
gives a methanol sputtering rate of $\sim 9\times10^{-21}\,\pers$ per
H atom or $\sim 2\times10^{-15}$ mol$^{-1}\,\pers$. 

By way of comparison, the direct cosmic ray-induced desorption rate of
water in conditions described above can be estimated to be
$\sim4.6\times10^{-21}\,\pers$ per H atom or
$\sim 9.3\times10^{-17}$\,mol$^{-1}\,\pers$. Here we have used the
sputtering flux from water ice derived by \cite{2019MNRAS.487.3392F},
$F^{\rm sp}_{\water} \sim f_{\water}\,
0.8\,\zeta^{\htwo}/10^{-17}\,\persqcm\,\pers$,
where $f_{\water}\sim 0.6$ is the fraction of water molecules in the
ice according to Table~2 of \cite{2015ARA&A..53..541B} (see also
Eq.~(7) of \citealt{2019MNRAS.487.3392F}). The fractional water
abundance relative to H atoms,
$n(\water,{\rm s})/n_{\rm H} \sim 5\times 10^{-5}$, is estimated in
the same manner as the fractional methanol abundance above.  Because
of similar binding energies of $\water$ and $\meth$, the desorption
rate estimate of \cite{2019MNRAS.487.3392F} should also be adequate
for methanol when corrected for the methanol fraction, $\sim0.08$ with
respect to $\water$. The resulting $\meth$ desorption rate
($\sim 3.7\times10^{-22}\,\pers$ per H atom or
$\sim 9.3\times10^{-17}$\,mol$^{-1}\,\pers$) is $\sim 60\%$ higher
than that derived using data from \cite{2019A&A...627A..55D}.  On the
other hand, applying similar scaling to the cosmic ray-induced
sputtering rate per water molecule,
$4.4\times10^{-17}$\,mol$^{-1}\,\pers$, derived by
\cite{2004ApJ...603..159B},
%and used later for example by \cite{2009ApJ...690.1497H} and
%\cite{2014MNRAS.440.2616K}
brings us very close to the result of \cite{2019A&A...627A..55D}, with
a methanol sputtering rate of $\sim 1.8\times10^{-22}\,\pers$ per H
atom or $4.4\times10^{-17}$\,mol$^{-1}\,\pers$. 
Finally, we note that the cosmic ray desorption rate per methanol
molecule used by \cite{1993MNRAS.261...83H} is very low compared to 
the values quoted above, 
$6.3\times10^{-20}$\,mol$^{-1}\,\pers$ ($\sim2.5\times10^{-25}\,\pers$
per H atom).

\subsection{Cosmic ray-induced photodesorption and dissociation}

\label{sec:photodesorption}

Photodissociation and photodesorption by the external
  far-ultraviolet (FUV) radiation field are considered to be important
  near cloud boundaries up to an efficient visual extinction of
  $A_V \sim 5^{\rm mag}$ (\citealt{2009ApJ...690.1497H};
  \citealt{2014MNRAS.440.2616K}). Stellar radiation in the Ophiuchus
  complex is exceptionally strong. \cite{2009ApJ...690.1497H} assume
  that the FUV radiation field in the vicinity of the core Oph\,A in
  the western part of the complex is 300 times the average local
  interstellar radiation field in that band as determined by
  \cite{1968BAN....19..421H} (often denoted by $G_0$).  Assuming that
  the FUV field is attenuated by 20 magnitudes in the ambient cloud
  (corresponding to an effective visual extinction of
  $A_V \sim 10^{\rm mag}$; e.g., \citealt{1989ApJ...345..245C}), the
  FUV flux at the core boundaries is $\sim 3\times 10^{-6} \,G_0$,
  that is, $\sim300$ FUV photons\,$\persqcm\,\pers$.  We model the FUV
  field and its attenuation in the case of H-MM1 in
  Section~\ref{sec:hmm1_illumination} below.

Cosmic ray protons and secondary electrons from cosmic ray ionization
sustain a flux of Lyman and Werner band photons ($11.2-13.6$ eV)
through the collisional excitation of $\htwo$ to excited 
electronic states \citep{1992MNRAS.258..125C}. According to the
estimates of \cite{1992MNRAS.258..125C} and
\cite{2018MNRAS.478.2753K}, the flux of cosmic ray-induced FUV photons
is $\sim 3000 - 5000\,\persqcm\,\pers$ in the interiors of dense dark
clouds. Often a flux value of $10^4\,\persqcm\,\pers$ or $10^{-4}\,G_0$ is used 
in chemistry models (e.g., \citealt{2014MNRAS.440.2616K}).   

The experimental results of \cite{2016ApJ...817L..12B} and
\cite{2016A&A...592A..68C} indicate that the efficiency of FUV
photodesorption of methanol is low, of the order of $10^{-5}$
molecules per incident photon. Assuming that dust grains are fully
covered by methanol ice, the photodesorption rate corresponding to
this efficiency is $\sim 3.3\times10^{-22}$ methanol
molecules\,$\pers$ per H atom. Using the composition of the ice
mantles in quiescent clouds and cores listed in Table~2 of
\cite{2015ARA&A..53..541B}, with $\sim5\%$ methanol, the
photodesorption rate becomes $\sim 1.6\times10^{-23}\,\pers$ per H
atom ($4.1\times10^{-18}$\,mol$^{-1}\,\pers$). Here we have adopted the
FUV flux $5000\,\persqcm\,\pers$ from \cite{1992MNRAS.258..125C} and
used the total surface area of grains ($4\times\sigma_{\rm H}$) from
the MRN size distribution.

\cite{2009A&A...504..891O} measured an average loss rate of condensed
methanol of $\sim 10^{-3}$/photon in irradiated ices. This loss rate
is probably attributable to photodissociation, producing fragments
such as ${\rm CH_3}$, ${\rm OH}$, and ${\rm CH_3O}$, that can remain
in the ice mantle or be ejected into the gas phase
(\citealt{2016ApJ...817L..12B}; \citealt{2016A&A...592A..68C}).  The
FUV photodissociation cross-section of methanol ice is
$\sim 2.7\times10^{-18}\,\sqcm$ (\citealt{2009A&A...504..891O};
\citealt{2016A&A...592A..68C}), so the photodissociation rate per
methanol molecule caused by the cosmic-ray induced FUV flux
$5000\,\persqcm\,\pers$ is $\sim 1.3\times10^{-14}$\,mol$^{-1}\,\pers$
(corresponding to $\sim 5.4\times10^{-20}\,\pers$ per H atom, assuming
$X(\meth,s)=4\times10^{-6}$).

Is cosmic ray-induced photodissociation of mantle species an important
source of chemical energy in dark cores? The ice mantle is supposed to
constitute $15-30\% $ of the grain mass in dense dark clouds, and it
is mainly composed of $\water$, $\co$, $\cotwo$, $\meth$, $\ammo$, and
$\methane$ \citep{2015ARA&A..53..541B}. Assuming that the mantle mass
is $15\%$ of that of the silicate core, one can calculate, adopting
the same FUV flux as above, that the number of photons impinging
onto a grain surface reaches $10\%$ of the number of mantle molecules
in $\sim 8,000$ yr.  The $10\%$ ratio of the number incident FUV
photons to the number of molecules was considered crucial in the
experiment of \cite{1982A&A...109L..12D}. The answer to the question asked 
above depends on the average number of heatings to the critical 
temperature 27\,K during the quoted period of time.    

Whole grain heating by cosmic rays has been recently modeled by
\cite{2018ApJS..239....6K} and \cite{2019MNRAS.486.2050K}.  Using
estimates of the energy spectrum of cosmic rays and their fluxes in
molecular clouds from \cite{2009A&A...501..619P} and
\cite{2016A&A...585A..15C}, Kalv{\=a}ns has derived cosmic ray heating
rates that (for moderate extinctions) are clearly higher than earlier
estimates (see Figure\,9 in \citealt{2018ApJS..239....6K}).  According
to this work, grains residing in an obscured cold cloud
($A_V\sim 10^{\rm mag}$) are heated to 27\,K every $\sim 900$ yr
(\citealt{2018ApJS..239....6K}; their Table~19).  This interval is
an order of magnitude shorter than the time needed for the accumulation of any
substantial amount of radicals through cosmic ray-induced photolysis.

\subsection{Reactive desorption}

\label{sec:reactive_desorption}

In reactive desorption, a molecule forming on a grain escapes into the
gas phase with the help of the exothermicity of the formation
reaction, without any external agent such as a cosmic ray or a photon
(e.g., \citealt{2006FaDi..133...51G}; \citealt{2007A&A...467.1103G};
\citealt{2013NatSR...3E1338D}; \citealt{2013ApJ...769...34V};
\citealt{2016A&A...585A..24M}).  The exothermicities of CO
hydrogenation reactions are of the order of a few eV, exceeding thus
clearly the binding energies of the products. The question is which
part of the released energy goes to the breaking of the chemical bond
to the surface and how much goes to the heating of the grain
(\citealt{2006FaDi..133...51G}). 

One can derive an estimate for the production rate of gaseous methanol
assuming that a certain fraction, $f_{\co}$, of CO molecules hitting grains
will be completely hydrogenated to methanol, and that a fraction, 
$f_{\meth}$, of methanol molecules so formed will be ejected into the gas. 
\cite{2006FaDi..133...51G} found that a low value of the
hydrogenation factor, $f_{\co}\sim0.003$, reproduces the observed
methanol abundances in their model. Another, much higher value of
$f_{\co} \sim 0.13$ can be derived based on the relative abundances
of carbon-bearing ice molecules in dark clouds ices listed in
\cite{2015ARA&A..53..541B}.

The fraction of desorbed products, $f_{\meth}$, also called the
efficiency of reactive desorption, has been found to depend strongly
on the substrate, so that it is much larger for a rigid surface than for
amorphous water ice (\citealt{2016A&A...585A..24M} and references
therein).  Based on the theoretical description of the reactive
desorption efficiency presented by \cite{2016A&A...585A..24M},
\cite{2017ApJ...842...33V} assumed that the efficiency of methanol
desorption upon formation on a CO rich surface is of the order of
$1\,\%$.  Similar desorption efficiencies have been used in previous
works (for example, \citealt{2007A&A...467.1103G} and
\citealt{2013ApJ...769...34V}).  According to the model of
\cite{2017ApJ...842...33V}, the formation of methanol is most
efficient in the ’CO freeze-out zone’ at densities of
$n(\htwo) \sim 10^4 - 10^5\,\percc$, where the gas-phase
  abundance of CO is still high, a few times $10^{-5}$ relative to
  $\htwo$. This model gives therefore an explanation for the fact
that $\meth$ seems to avoid the densest parts, and sometimes shows a
shell-like distribution (\citealt{2006A&A...455..577T};
\citealt{2014A&A...569A..27B}; \citealt{2018ApJ...855..112P}).

The density of CO molecules in the CO-freezing zone is similar to
  those of H atoms, that is $n_{\co}\sim 1\,\percc$
  ($X(\co)=10^{-5}-10^{-4}$) but their average thermal speed is a
  factor of 5 lower, ${\bar v}_{\co}\sim 8,700\,\cms$.  Using the
  cross-sectional area of grains from the MRN size distribution, and
  the $\co$ flux implied by the density and the speed quoted above,
  the methanol production rate by reactive desorption is
  $\sim 4.3\times10^{-22}$ or $1.9\times10^{-20}$ molecules\,$\pers$
  per H atom ($1.1\times10^{-16}$ or
  $4.6\times10^{-15}$\,mol$^{-1}\,\pers$), depending on whether we
  adopt $f_{\co}=0.003$ or $f_{\co}=0.13$. The higher of these rates
  is comparable to the cosmic ray-induced sputtering rate of methanol
  from CO ice estimated in Section~\ref{sec:cr_desorption}, based on
  the results of \cite{2019A&A...627A..55D}.

\subsection{Grain-grain collisions induced by turbulence}

\label{sec:turbulence}

As discussed by \cite{1982A&A...109L..12D}, low-velocity collisions
between grains can lead to grain heating above the temperature
threshold ($\sim30$\,K) that triggers explosive radical-radical
reactions and the partial disruption of the grain mantle. Assuming
that the grains consist of a silicate core and a mantle of water ice
constituting $\sim 15\%$ of the grain mass, one finds that the enthalpy change
needed to raise the grain temperature from 10\,K to 27\,K corresponds
to a collision velocity of $\sim30\,\ms$. Here we have adopted the
heat capacity functions from \cite{1985A&A...144..147L} (their Eq.~1)
and \cite{2004A&A...416..187S} (their Eq.\,4) for the grain core and
the ice mantle, respectively. The collisional speed that is needed 
to heat these grains to 100\,K is $180\,\ms$.
 
Collisions between grains require that the grain population has
acquired an internal velocity dispersion. This can arise when grains
are embedded in a turbulent gas (\citealt{1978M&P....19..221V};
\citealt{1985prpl.conf..621D}).  In the turbulent acceleration model
presented by \cite{1985prpl.conf..621D}, the grain velocities are
determined by turbulent velocity fluctuations occurring on a time
scale comparable to that of the hydrodynamical drag. In Kolmogorov
turbulence, the velocity is proportional to the square-root of the
eddy turn-over time. Because the drag time is directly proportional to
the grain radius, $a$, the velocity distribution of large grains has a
square-root dependence on the grain radius, $v \propto a^{1/2}$.  The
smallest grains are also coupled to the magnetic 
field\footnote{A great majority of grains is negatively charged at
  visual extinctions above $A_{\rm V} \sim 3^{\rm mag}$
  \citep{2015ApJ...812..135I}.},
and their velocity dispersion is determined by turbulent fluctuations
on the time scale comparable to the Larmor time
(\citealt{2002ApJ...566L.105L}; \citealt{2004ApJ...616..895Y}).  This
causes a $v \propto a^{3/2}$ dependence for the smallest grains.
Overall, turbulent acceleration is more effective for large grains
than for small grains. This gives rise to velocity differences between
small and large grains, and to an enhanced rate of grain-grain
collisions.

\begin{figure}[]
\unitlength=1mm
\begin{picture}(80,61)(0,0)
\put(0,0){
\begin{picture}(0,0) 
\includegraphics[width=8cm,angle=0]{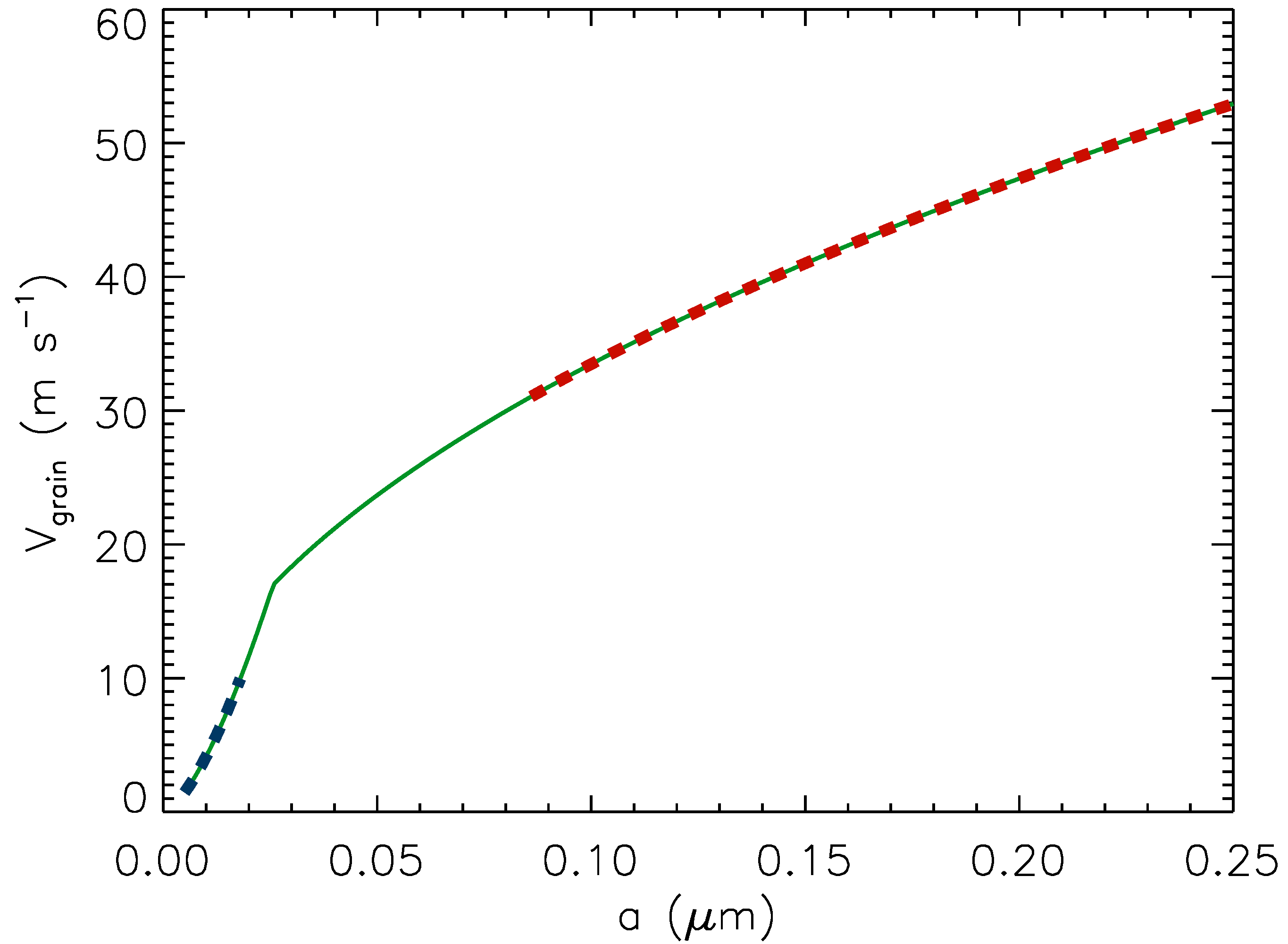} 
\end{picture}}
\end{picture}
\caption{Relative speed of grains as a function of the grain radius
  according to the turbulent acceleration model
  (\citealt{1985prpl.conf..621D}; \citealt{2002ApJ...566L.105L}) in
  dense dark cloud conditions (see text). The regimes of ``high'' and
  ``low'' velocity grains are indicated with thick dashed lines
    in red and blue, respectively. The discontinuity in the gradient
  is caused by the coupling of the smallest grains to the magnetic
  field. This occurs at $a=0.026\,\mum$.}
\label{figure:vgrain}
\end{figure}

In Figure~\ref{figure:vgrain} we show relative grain velocities as a
function of grain radius according to the models of
\cite{1985prpl.conf..621D} and \cite{2002ApJ...566L.105L}, for
conditions characteristic of the outer envelope of a cold dense core, 
$n(\htwo)=10^5$\,cm$^{-3}$, $T=10$\,K, $B=100\,\mu$G. Here it is
assumed that the turbulent velocity field has a Kolmogorov-like
spectrum, $v \propto l^{1/3}$. The absolute scale is set by assuming
that the turbulent velocity is $0.7\,\kms$ on the scale of 0.034\,pc. 
This corresponds to the velocity gradient across the H-MM1 core 
(Sect.~\ref{sec:line_analysis}; for discussion about
the connection between velocity gradients and turbulence see
\citealt{2000ApJ...543..822B}).  Assuming the MRN grain size
distribution, the small, slow grains with
velocities below $10\,{\ms}$ comprise $\sim 50\%$ of the total surface
area of the grains, whereas the share of large, fast grains with
$v > 30\,\ms$ is only $\sim 5\%$ of the surface area (albeit $\sim 50\%$ of the dust mass). 

The frequency at which a ``slow'' grain collides with a ``fast'' grain
is obtained from
\begin{equation}
f_{\rm coll} = n_{\rm g}^{\rm fast} \; 
\pi ({\bar a}^{\rm fast} + {\bar a}^{\rm slow})^2 \, 
( {\bar v}_{\rm g}^{\rm fast} - {\bar v}_{\rm g}^{\rm slow} ) \; ,
\end{equation}
where $n_{\rm g}^{\rm fast}$ is the number density of fast grains,
${\bar a}^{\rm fast}$ and ${\bar a}^{\rm slow}$ are the average radii
of the fast and slow grains, respectively,
${\bar v}_{\rm g}^{\rm fast}$ and ${\bar v}_{\rm g}^{\rm slow}$ are
their average speeds.  Here we have used the classical
  collisional cross-section for hard spheres (e.g.,
  \citealt{1997A&A...322..296C}). The MRN grain size distribution and
  the physical parameters mentioned above imply the following values
  for the average radii and speeds:
  ${\bar a}^{\rm slow}\sim 80$\,{\AA},
  ${\bar a}^{\rm fast}\sim 0.12\,\mum$,
  ${\bar v}_{\rm g}^{\rm fast}-{\bar v}_{\rm g}^{\rm slow} \sim 34\,\ms$.  The number density of   ``fast'' grains is
  $n_{\rm g}^{\rm fast} \sim 8\times10^{-8}\, \percc$. With these
  parameters, the collision frequency is very low,
  $f_{\rm coll} \sim 1.4\times10^{-13}\,\pers$, meaning that a slow
  grain collides with a fast one once in $\sim 200,000$\,yr. Using the actual
  grain size and speed distributions without averaging, one obtains
  $\sim 25\%$ higher grain-grain collision rate at a minimum speed
  difference of $30\,\ms$. 

When both particles are negatively charged, the kinetic energy
  overcomes the Coulomb barrier at this collision speed. We assume
  that a grain carries maximally one electron charge. The
  electrostatic repulsion can be important in collisions between the
  smallest grains with steep Coulomb potentials and small velocity
  fluctuations (\citealt{1993ApJ...407..806C};
  \citealt{2011ApJ...727....2P}). In high-speed collisions, one of the
  grains is large, $a\sim 0.1\,\mum = 10^{-5}$\,cm, and the minimum
  collision speed needed to overcome the Coulomb barrier is probably
  lower than the critical speed for sticking
  (\citealt{1993ApJ...407..806C}, Fig.~8).

Despite the low collision frequency, the desorption rate can be
  high if every collision between a slow and a fast grain lead to a
  chemical explosion and the complete disruption of the ice mantle of
  the smaller, slow grain. In this case the desorption rate of methanol would be
  $f_{\rm coll} \, n_{\meth}^{\rm surf, slow}$ molecules
  $\percc\,\pers$, where $n_{\meth}^{\rm surf, slow}$ is the number
  density of methanol molecules residing on slow grains (corresponding
  in our example to $\sim 50\%$ of their total number density).
  It is, however, questionable if mild grain-grain collisions that
  heat small grains to $\sim 27$\,K can trigger mantle explosions,
  when these grains are, according to the results of
  \cite{2018ApJS..239....6K}, heated to the same temperature much more
  often by cosmic rays (see end of
  Section~\ref{sec:photodesorption}), and the accumulation of radicals 
 in the mantle is thereby inhibited.

 Therefore it seems plausible that only energetic grain-grain
 collisions that heat them close to $\sim 100$\,K can lead to
 significant desorption of methanol through thermal evaporation.  The
 maximum grain speed acquired in Kolmogorov turbulence depends on the
 density and temperature according to
 $v_{\rm max} \propto n^{-1/2}\,T^{-1/4}$
 \citep{1985prpl.conf..621D}. At densities above
 $\sim 3\times 10^5\,\percc$ the maximum speed drops below the
 critical value $30\,\ms$.  Assuming that the temperature remains
 constant at 10\,K, the maximum densities where grain-grain collisions
 are energetic enough to heat them to 100\,K (with a collisional speed
 of $\sim 180\,\ms$) is $n(\htwo) \sim 8\times 10^3\,\percc$.
 Grain-grain collision frequencies decrease toward lower densities, as
 is likely to happen to the methanol abundance on grains. 
 
 Following \cite{2004A&A...415..203S} (see their Eq. 29), we assume
 that collisional energy in excess of heating a grain to 26\,K is
 available for sublimation.  In a totally inleastic collision at
 $180\,\ms$ this excess energy per unit mass is
 $\sim 16$\,J\,g$^{-1}$.  Adopting the mantle composition from
 \cite{2015ARA&A..53..541B}, and assuming that the mantle constitutes
 15\% of the grain mass, the total binding energy of the mantle
 molecules per one gram of dust is $\sim 230$\,J\,g$^{-1}$. Here we
 have used the following binding energies: $\water$ 5700\,K, CO
 1150\,K, CO$_2$ 2600\,K, $\meth$ 5500\,K, $\ammo$ 5500\,K, CH$_4$
 1300\,K. So it seems that the energy deposited by a grain-grain
 collision at $180\,\ms$ can evaporate at most $\sim 7\%$ of the
 mantle material. The minimum collision speed required for the
 evaporation of the whole mantle is $\sim 680\,\ms$. Neglecting the
 differences in the binding energies, that is, assuming that the
 mantle species are evaporated in proportion to their relative
 abundances, we obtain a methanol desorption rate of
 $\sim 4\times10^{-21}\,\pers$ per H atom or
 $\sim 1\times10^{-15}$\,mol$^{-1}\,\pers$ at the density
 $n(\htwo) \sim 8\times 10^3\,\percc$ (by collisions at the minimum
 speed of $180\,\ms$). However, the presence of SO emission in the
 same region where we see $\meth$ lines in H-MM1 suggests that the
 density methanol desorption layer is of the order of
 $\sim 10^5\,\percc$ (see Section~\ref{sec:hmm1}). Kolmogorov
 turbulence at this density is unlikely to induce collisions leading
 to signifant thermal evaporation.

\subsection{Grain-grain collisions in a shear layer}

\label{sec:shear}

The last mechanism discussed here is the release of methanol in
grain-grain collisions induced by shear instability. This is motivated
by the fact that the methanol distribution observed in H-MM1 
(Figure~\ref{figure:xmeth}) resembles Kelvin-Helmholtz clouds that
sometimes develop in the atmosphere in flows with large vertical
shears. 
As described in standard textbooks, such as \cite{Batchelor1967},
advection of vorticity amplifies the dominant sinusoidal disturbance
in a vortex sheet and makes it roll up. The evolution of vortex sheets
has been studied numerically and semi-analytically by
\cite{1976JFM....73..215P} and \cite{1976JFM....73..241C}. According
to their results, vorticity reaches its maximum values at places they
call ``braids'' (near the troughs) and ``cores'' or cat's eyes (near
the crests), separated by approximately half the wavelength of the
dominant disturbance.  

We conjecture that vorticity in the shear layer can accelerate dust
grains to velocities deviating from the mean flow, and induce
grain-grain collisions, in the same manner as Kolmogorov-type
turbulence discussed in the previous section. However, the energy
spectrum and the corresponding velocity scaling law in a vortex sheet 
are different from those in fully developed turbulence 
(\citealt{1967PhFl...10.1417K}; \citealt{1971JFM....47..525K}; 
\citealt{1988JFM...193..475G}; \citealt{2011PhRvE..84b6318A}). 

The time-scale of velocity fuctuations depends on the free-stream
velocity and the size-scale of vorticity. The drag time for
$0.1\,\mum$ grains is $\sim 100$\,yr at
$n(\htwo)\sim10^5\,\percc$. Assuming that the free-stream velocity is 
transonic, $U\sim 200\,\ms$, one obtains that only small vortices of size
$\lambda \la 5$\,au can give rise to significant velocity differences
amongst the grain population.

\begin{figure}[]
\unitlength=1mm
\begin{picture}(80,70)(0,0)
\put(-5,-2){
\begin{picture}(0,0) 
\includegraphics[width=9cm,angle=0]{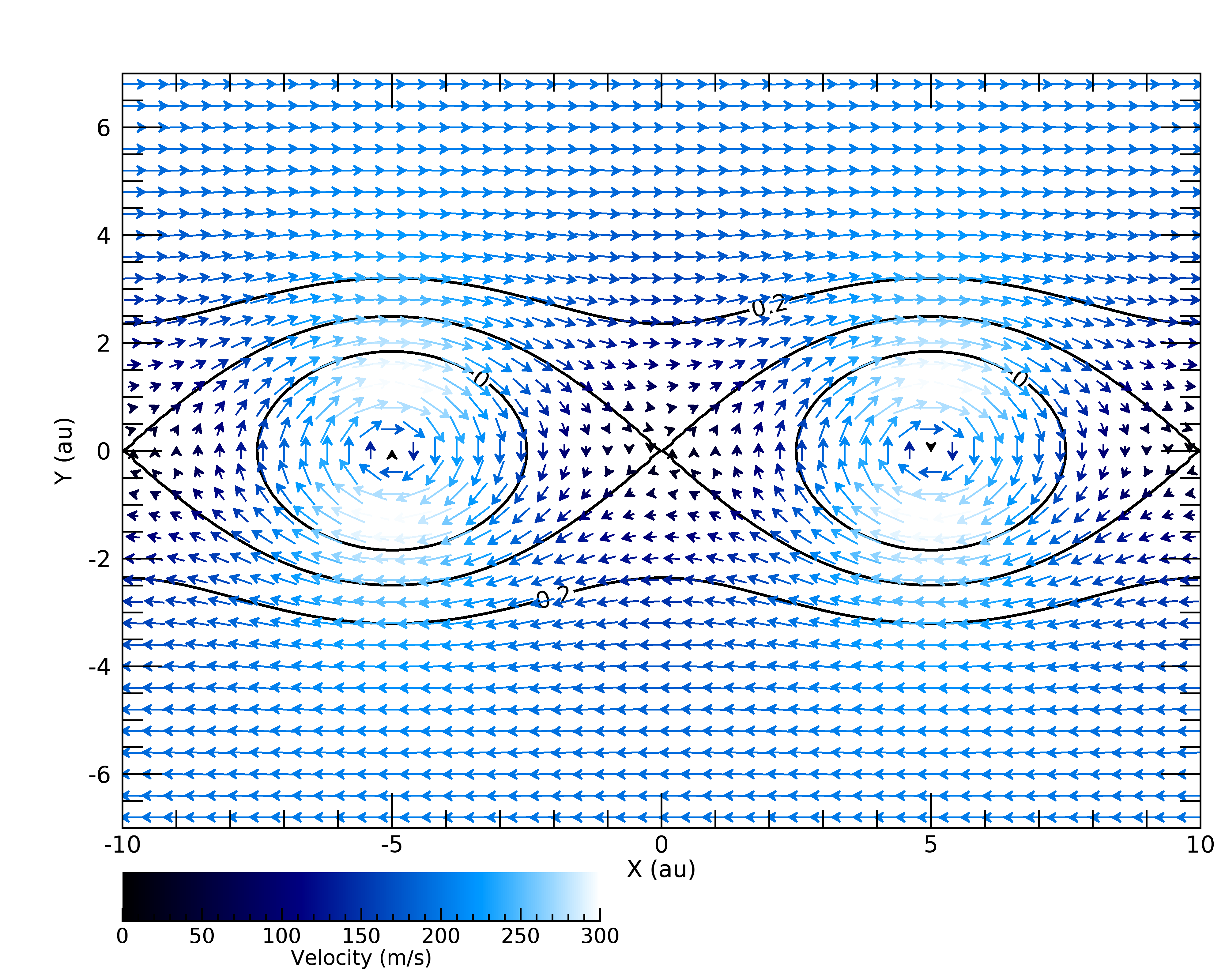} 
\end{picture}}
\end{picture}
\caption{Velocity field in a periodic inviscid vortex sheet according to \cite{1976JFM....73..241C}. Some of the streamlines, corresponding to the 
stream function values $\Psi=0$, 0.1, and 0.2 times the maximum, are shown. The cat's eyes are outlined by the contour 0.1.}
\label{figure:vortex_sheet}
\end{figure}

In Figure~\ref{figure:vortex_sheet} we plot the velocity field
calculated from the stream function given by
\cite{1976JFM....73..241C} (their Eq.~3.9). Some of the streamlines
are also plotted, including the so called stagnation streamline which
outlines the cat's eyes. In this solution, the traverse time is not
constant inside the cat's eyes, but increases steeply from the centre,
and reaches the maximum ($\sim 300$\,yr in this example with
$U=200\,\ms$, $\lambda=10$\,au) on the stagnation streamline. Beyond
this streamline, the period of the smooth oscillatory motions attains
quickly a constant value ($\sim 230$\,yr).

In the structure shown in Figure~\ref{figure:vortex_sheet}, the
turn-over time of the flow is shorter than the drag time of
$0.1\,\mum$ grains within the contour $\Psi = 0$. One could expect
grain-grain collisions to be concentrated to these regions, where
small grains should be advected by the swirling flow, but large grains
will not respond to the velocity fluctuations. For still smaller
vortices, stagnation points between the cat's eyes (at the origin of
Figure~\ref{figure:vortex_sheet}) are likely places of vigorous
grain-grain collisions. Because of the high speed of the flow, the
collisions are more frequent and more energetic than those induced by
Kolmogorov turbulence. Assuming an average collision speed of
  $200\,\ms$ and making the division between small and large grains at
  $a=0.1\,\mum$, a small grain collides at an average frequency of
  $\sim7\times10^{-13}\,\pers$ (every 45,000\,yr) at a density of
  $n(\htwo)\sim 10^5\,\percc$. At this collision speed, the excess
  energy after heating the grains to 26\,K is $\sim 20$\,J\,g$^{-1}$,
  and we assume that this is sufficient for evaporating $9\%$ of the
  mantle material (see end of Section~\ref{sec:turbulence}). 

  The resulting methanol desorption rate is $\sim 3\times 10^{-14}$
  molecules $\percc\,\pers$ or $\sim 2\times 10^{-19}\,\pers$ per H
  atom, corresponding to $\sim 4\times10^{-15}$\,mol$^{-1}\,\pers$.
  Here we have again assumed that the fractional abundance of methanol
  ice is $4\times10^{-6}$ with respect to H atoms. The desorption rate
  by shear-induced grain-grain collisions is comparable to the rates
  of reactive desorption and cosmic-ray induced sputtering in their
  upper ranges. This mechanics is, however, confined to very small
  volumes near discontinuities in the velocity field. Assuming that
  the vortex sheet is linear with a thickness of 5\,au, it fills
  approximately 1\% of the 560\,au beam of the present observations. A
  winding structure would increase the beam filling factor
  slightly. Nevertheless, taking the non-uniform beam filling into
  account, the contribution of grain-grain collisions in shear layers
  to the observed methanol abundance can be locally similar to those
  of the more widespread reactive desorption and cosmic-ray induced
  sputtering. The estimated methanol desorption rates for different
  mechanisms are summarized in Table~\ref{table:desrates}. 

\begin{table}[htb]
\caption{Estimated $\meth$ desorption rates and the sections where they 
are discussed.}
\label{table:desrates}
{\begin{center}
\small
\begin{tabular}{lcl}
% &  \\
Mechanism & $k_{\rm des}$ (mol$^{-1}\,\pers$) & Sect. \\ \hline
cosmic-ray desorption &  $6\times10^{-17} - 2\times10^{-15}$ & 
\ref{sec:cr_desorption} \\
photo-desorption & $4\times10^{-18} - 8\times10^{-17}$ &  \ref{sec:photodesorption} \\
reactive desorption & $1\times10^{-16} - 5\times10^{-15}$ & 
\ref{sec:reactive_desorption} \\
grain-grain collisions &  & \\
Kolmogorov turbulence &  $\sim 1\times10^{-15}\,^{a}$ & \ref{sec:turbulence} \\
shear instability &  $\sim 4\times10^{-15}\,^{b}$ & \ref{sec:shear} \\ \hline
\multicolumn{3}{l}{$^a$ unlikely to work at high densities} \\
\multicolumn{3}{l}{$^b$ localized to small areas} \\
\end{tabular}
\end{center}
}   
\end{table}

\section{Origin of gaseous methanol in H-MM1}

\label{sec:hmm1}

In accordance with previous mappings toward L544
  (\citealt{2014A&A...569A..27B}; \citealt{2017A&A...606A..82S};
  \citealt{2018ApJ...855..112P}), we find relatively high abundances
of methanol in a prestellar core, offset from the density maximum.
With ALMA, the methanol distribution in H-MM1 could be mapped
  with a higher spatial resolution than achieved in L1544, and it was
found to be confined to a layer that follows the core boundaries. The
apparent thickness of this layer is of the order of 1000\,au,
  except for the bright curved feature at the southeastern boundary
  that is unresolved with our 560\,au beam.  The critical density of
the coexistent SO$(3_2-2_1)$ line is $3\times 10^5\,\percc$ at 10\,K,
whereas for the methanol lines this is $3\times10^ 4\,\percc$. We
assume that the higher of these values, $\sim 10^5\,\percc$, is
characteristic of the gas component detected in $\meth$ and SO. This
value agrees with the densities derived by \cite{2016A&A...587A.130B}
for methanol emission regions in prestellar cores.  At still higher
densities, traced by $\dammo$, both $\meth$ and SO are apparently
frozen out.

Sulphur monoxide can form both in the gas phase and on grain
surfaces. Without efficient desorption, gaseous SO depletes quickly at
high densities. On the other hand, as soon as atomic sulphur is
available, SO and other sulphur-bearing molecules are thought to form
quickly in the gas phase through reactions with O, $\otwo$ and OH
(\citealt{2016A&A...593A..94F}; \citealt{2017MNRAS.469..435V}). 
  According to modeling results of \cite{2019A&A...624A.108L}, the
  principal sulphur-bearing species in the ice at advanced chemical
  stages are SO, OCS, HSO, and CS. Also pure-sulfur molecules, 
in particular S$_8$, have been suggested to be abundant on grains 
\citep{2020ApJ...888...52S}.  Methanol and sulphur monoxide are
not directly related. Their spatial coincidence could possibly be
explained by codesorption of $\meth$ and SO or sulphur in some other
form. Alternatively, gas-phase reactions forming SO may be efficient
in the same conditions where $\meth$ is desorbed.

According to the estimates presented in the previous section, the
efficiencies of reactive desorption and desorption in grain-grain
collisions in the presence of vigorous velocity fluctuations can be
comparable to the efficiency of direct cosmic ray-induced desorption,
but the two mechanisms mentioned first are likely to be spatially
limited to certain zones of a cloud.  The reactive desorption model
predicts that the $\meth$ and SO abundances peak on the outskirts of a
dense core (\citealt{2017ApJ...842...33V}, their Figures~4 and 8). In
this model, gaseous methanol is abundant at a depth where water is
partially photodissociated (Figure~6 of \citealt{2017ApJ...842...33V}).
The presence of hydrogen atoms and hydroxyl radicals in this layer can
also promote the formation of SO in the gas phase.  On the other hand,
if methanol formation is most efficient in the CO-freezing zone, a
halo-like structure does not depend on the actual desorption
mechanism. Also cosmic-ray induced desorption can be thought to give
rise to a similar distribution, especially if one takes the reduction
of cosmic ray flux toward the interior parts into account
(\citealt{2009A&A...501..619P}; \citealt{2015ApJ...805...59I}).
 
The strong asymmetry observed in H-MM1 is, however, difficult to
explain without invoking an external agent which either destroys
$\meth$ and SO on one side of the core or creates them on the opposite
side through forceful desorption.  We first discuss the possible effect
of an asymmetric radiation field.

\subsection{Uneven illumination}

\label{sec:hmm1_illumination}

An asymmetric methanol distribution has been previously found in the
prestellar core L1544 (\citealt{2016A&A...592L..11S};
\citealt{2018ApJ...855..112P}).  \cite{2016A&A...592L..11S} suggested
that the methanol distribution in L1544 reflects asymmetric
illumination which hinders CO production on the more exposed side of
the cloud. Intense radiation could also inhibit SO formation by 
  ionization. The ionization potentials of SO and the S atom are
  relatively low, 10.29\,eV and 10.36\,eV, respectively; the limiting
  wavelength of ionizing radiation, $\lambda < 120$\,nm, falls in the
  range where photons also can dissociate $\meth$ efficiently
  ($114\,{\rm nm} < \lambda < 180\,{\rm nm}$;
  \citealt{2016A&A...592A..68C}).

  The ambient cloud around H-MM1 continues towards the east, giving a
  reason to believe that the interstellar radiation field is stronger
  on the western side where $\meth$ and SO are weaker.  This side
  faces the Upper Sco-Cen (USC) subgroup of the Scorpius-Centaurus OB
  association, including the luminous B-type double stars $\rho$ Oph
  and HD 147889 (\citealt{1992A&A...262..258D};
  \citealt{2008hsf2.book..235P}; \citealt{2019A&A...623A.112D}).  The
  blue subgiant binary HD 147889, about 1.2\,pc west of H-MM1, is the
  dominant UV source in the region (\citealt{1999A&A...344..342L};
  \citealt{2008MNRAS.391.1075C}; \citealt{2013MNRAS.428.2617R}).

If the radiation field is stronger on the western side, this should be
evident as a dust temperature difference between the two sides of the
core because dust grains are primarily heated by absorption of
starlight. We examined this hypothesis using mid- to far-infrared maps
from Spitzer, Herschel, and SCUBA-2, and found that the dust
temperature reaches, indeed, its minimum on the eastern side of the
core, close to the $\meth$ maximum. The analysis of the dust continuum
maps is presented in Appendix~\ref{sec:tdust}. The analysis suggests a
temperature drop from about $14-15$\,K on the western edge to about
$11-12$\,K in the eastern cove of the core (see
Fig.~\ref{figure:taumidfar}).

We tested this result by simulating the effect of an asymmetric
  radiation field on a core model resembling H-MM1. The strength of
  the external field was adjusted until the model could reproduce the
  observed surface brightness maps at $850\,\mum$.  The
  details of the calculation are presented in
  Appendix~\ref{sec:tdust}. The simulation shows that an anisotropic
  radiation field causes a displacement of the temperature minimum
  from the density maximum, and that this displacement can be inferred
  by comparing $70$ and $850\,\mum$ surface brightness maps. It also
  gives estimates for the strength of the FUV radiation field on the eastern
  and western edges of the core. The energy density of the FUV field
  along three cuts accross the model core is shown in
  Figure~\ref{figure:radiation_field}. The strongest methanol emission
  in H-MM1 is found at the offsets $10\arcsec-20\arcsec$ east of the
  density maximum (the map center). In the simulated core, the FUV
  energy density has a deep minimum in this range, and the average
  radiation flux originating from the external field is negligible. In
  the corresponding region on the western side of the core, the
  average FUV flux is approximately $F_{\rm FUV}=80,000$
  photons\,$\persqcm\,\pers$, that is, an order of magnitude higher
  than the flux of cosmic ray-induced photons ($F_{\rm
    FUV}=5,000\,\persqcm\,\pers$) in dense dark clouds estimated by
  \cite{1992MNRAS.258..125C}.

The difference in the FUV fluxes at the western and eastern
boundaries of the core is likely to affect the methanol abundances
both in the gas and on grains. Assuming that $\meth$ is formed on grains
through CO accretion, and removed by photodissociation and by some generic desorption mechanism, its equilibrium abundance on grains relative to H atoms can be written as
\begin{equation}
X(\meth,{\rm s}) = \frac{f_{\co}\, \sigma_{\rm H} \, {\bar v}_{\co} \, n_{\co}}
{\sigma_{\rm ph,s}\,F_{\rm FUV} + k_{\rm des}} \; , 
\label{eq:xgrain}
\end{equation}
where $f_{\co}$ is the efficiency factor for the conversion $\co \rightarrow \meth$ on grains, ${\bar v}_{\co}$ is the average thermal speed of CO molecules,  $n_{\co}$ is their number density, $\sigma_{\rm ph,s}=2.7\times10^{-18}\,\sqcm$ is the FUV photodissociation cross-section of methanol ice
\citep{2016A&A...592A..68C}, and $k_{\rm des}$ is the desorption rate
per molecule $\pers$. We assume here that the desorption rate is
constant and represents the high end of the various estimates above,
that is, $k_{\rm des} = 5\times10^{-15}$ mol$^{-1}\,\pers$. For simplicity, we also assume the number density and the average speed of CO molecules are 
constant in the core, $n_{\co}=1\,\percc$, ${\bar v}_{\co} = 8,700\,\cms$.
Moreover, we assume that a constant cosmic-ray induced FUV flux,
$5,000\,\persqcm\,\pers$ comes everywhere on top of the attenuated
external field. We then adjust the efficiency factor $f_{\co}$ so that the
maximum fractional methanol abundance on grains is
$4\times10^{-6}$. The resulting factor is $f_{\co}=0.05$.

\begin{figure}[]
\unitlength=1mm
\begin{picture}(80,61)(0,0)
\put(0,0){
\begin{picture}(0,0) 
\includegraphics[width=8cm,angle=0]{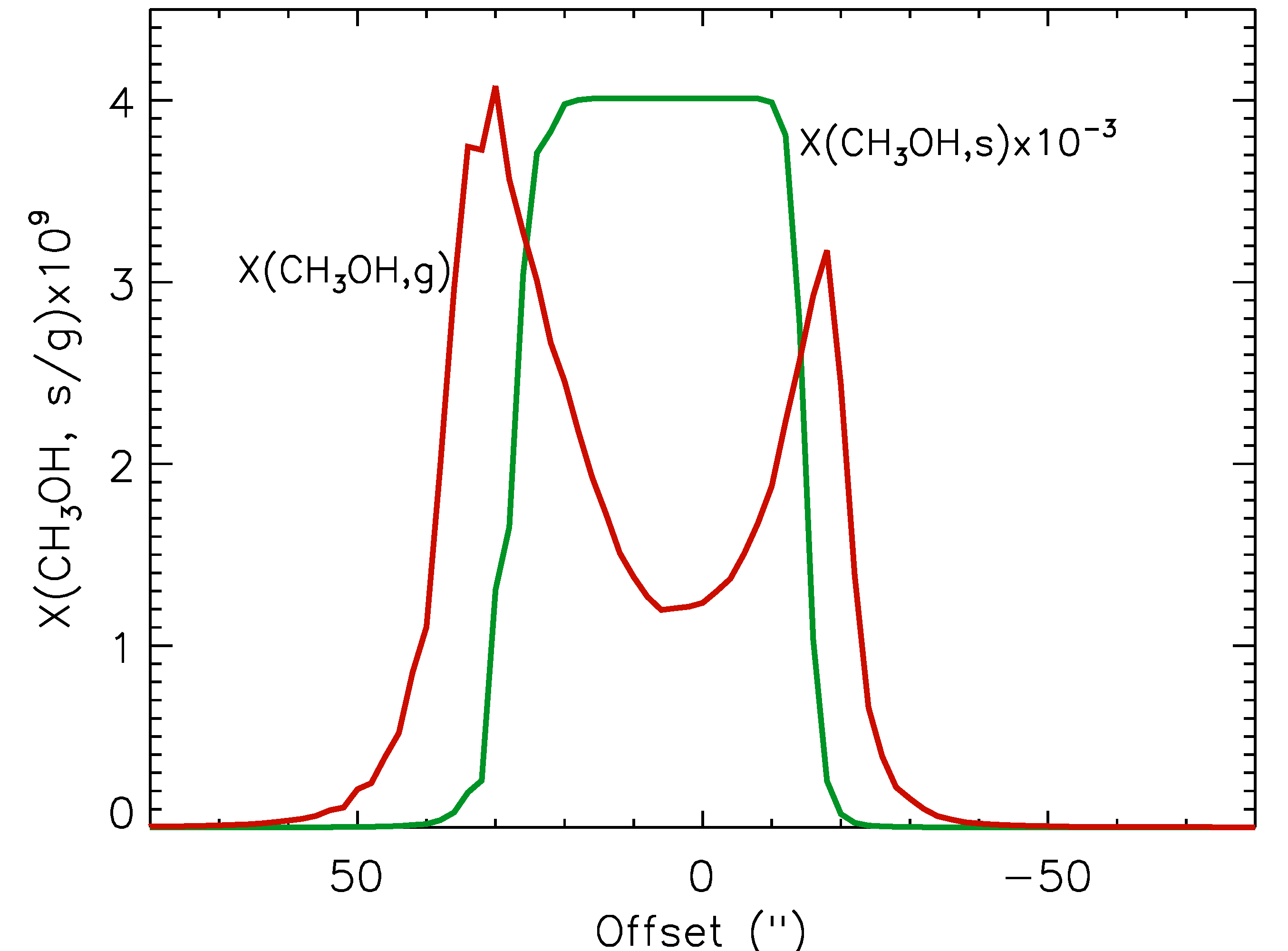} 
\end{picture}}
\end{picture}
\caption{Fractional methanol abundances on grain (s) and in the gas
  (g) relative to H atoms as functions of distance from the core
  center. The abundances are calculated using Eqs.~\ref{eq:xgrain} and
  \ref{eq:xgas} together with the FUV flux and particle density from
  our 3-dimensional model of H-MM1. The offsets are measured along a
  line lying in the plane of the sky.}
\label{figure:xsg_meth}
\end{figure}

Assuming that accretion onto grains and photodissociation dominate the
destruction of gas-phase methanol, its equilibrium abundance can be
obtained from
\begin{equation}
X(\meth,{\rm g}) = \frac{X(\meth,s)\, k_{\rm des}}
{n_{\rm H}\,\sigma_{\rm H}\,{\bar{v}}_{\meth} + \sigma_{\rm ph,g}\,F_{\rm FUV}} \; ,
\label{eq:xgas}
\end{equation}
where ${\bar{v}}_{\meth}$ is the average thermal speed of methanol
molecules ($\sim 8,100\,\cms$ at 10\,K), and $\sigma_{\rm ph,g}$ is
the FUV photodissociation cross-section of methanol in the gas. For
the latter we adopt the value
$\sigma_{\rm ph,g}\sim 1\times10^{-17}\,\sqcm$ based on the
experimental results of \cite{1971JChPh..55.3390P} and
\cite{2002JChPh.117.1633C}. The hydrogen density, $n_{\rm H}$, comes
from our core model.  The grain-surface and gas-phase abundances of
methanol calculated from Eqs.~\ref{eq:xgrain} and \ref{eq:xgas} along
a horisontal line that goes through the density maximum of the core
model are plotted in Figure~\ref{figure:xsg_meth}. The abundance
profile in the gas phase is double-peaked, with the stronger peak
residing on the dark side of the core, whereas on grains the abundance
profile is flat-topped. The eastern and western gas-phase peaks, with
$X(\meth,g)\sim 4\times10^{-9}$ and $X(\meth,g)\sim 3\times10^{-9}$,
occur at the densities $n(\htwo)\sim 1.8\times10^{5}\,\percc$ (east)
and $n(\htwo)\sim 2.0\times10^{5}\,\percc$ (west). The fractional
grain-surface methanol abundances at these locations are
$X(\meth,s)\sim 1.3\times10^{-6}$ (east) and
$X(\meth,s)\sim 2.6\times10^{-7}$ (west). The stronger methanol peak
lies much further out from the core center than in the observed core,
but we cannot exclude the possibility that the difference is caused by a
projection effect.

This simple model can be envisaged to give a qualitative explanation
for the methanol distribution observed in H-MM1, even though
Figure~\ref{figure:xsg_meth} shows the methanol abundance along a line
lying in the plane of the sky, while we are measuring line-of-sight
averages.  However, the contrast between the peak methanol abundances on the
eastern and western sides, $\ga10:1$, cannot be reproduced by this
model. Increasing the external FUV flux would shift the methanol peaks
inward, toward higher densities, but the abundance ratio would remain
the same, approximately $4:3$. Likewise, one can increase or decrease 
the gas-phase abundances by adjusting the desorption rate with the same effect
on the abundance ratio.

Besides indicating asymmetry in the radiation field, the dust
temperature drop itself can contribute to the high abundances of
$\meth$ on the eastern side. Methanol production is probably
favoured by efficient accretion and hydrogenation of CO on grains at
low temperatures. It should be noted, however, that the cooler region
does not cover the whole eastern side of the core where strong $\meth$
emission is found, and it does not extend to the SO maximum. At the
northern and southern ends of the integral-shaped $\meth$ and SO
emission region, the dust temperature is similar to that on the western
side where these species show only weak emission.

The estimates and considerations presented above suggest that in
addition to the fact that the H-MM1 core is exposed to a strong,
asymmetric radiation field that hinders methanol formation on its
western side, desorption is particularly effective at its eastern
boundary.  Of the mechanisms discussed in
Section~\ref{sec:desorption}, photodesorption is very unlikely to
cause this effect. Also, it is not likely that the cosmic ray flux
shows substantial anisotropy on this scale, or that the efficiency of
reactive desorption is higher in the east, except perhaps in the
vicinity of the temperature minimum (Appendix~\ref{sec:tdust}). In
what follows, we therefore consider desorption related to gas dynamics
at the core boundaries.

\subsection{Shocks}
\label{sec:hmm1_shocks}

Both $\meth$ and SO are known to increase in shocks, and they have
been used to probe outflows and the accretion process associated with
star formation (\citealt{1997ApJ...487L..93B};
\citealt{2015A&A...581A..85P}; \citealt{2016ApJ...824...88O}). Strong
enhancement of SO and $\sotwo$ is predicted by models of magnetized
molecular C-shocks, as a result of neutral-neutral reactions in the
shock-heated gas and the erosion of S-rich icy grain mantles owing to
bombardment by heated gas particles (\citealt{1993MNRAS.262..915P};
\citealt{1994MNRAS.268..724F}). Sputtering of the grain mantles
associated with shocks can also increase the $\meth$ abundance
substantially in the gas phase \citep{2008A&A...482..549J}.  Previous
observations of mid-$J$ CO rotational lines toward Perseus and Taurus
complexes suggest low-velocity shocks associated with dissipation of
turbulence and core formation in molecular clouds
(\citealt{2014MNRAS.445.1508P}; \citealt{2015ApJ...806...70L}).
Judging from the fact that the non-thermal velocity dispersion
experiences an abrupt change at the core boundary
(\citealt{2019ApJ...872..207A}; see Section~\ref{sec:hmm1_turbulence}),
low-velocity shocks caused by accreting material are also possible in the
case of H-MM1. However, the present data consisting of low-lying
rotational lines $\meth$, SO, and $\dammo$ do not show any evidence of
shock heating or velocity gradients that would be large enough
($\Delta v \ga 10\,\kms$) to give rise to significant shock-induced
sputtering (\citealt{1997A&A...322..296C}; \citealt{2008A&A...482..549J}).

\subsection{Desorption in the turbulent envelope}

\label{sec:hmm1_turbulence}

The interior parts of the H-MM1 core have a very low level of
non-thermal motions. This is evident from the velocity dispersion of
the ortho-$\dammo$ lines illustrated in Figures~\ref{figure:sigma_maps}a and
\ref{figure:sigma_histos}).  In a study based on $\ammo(1,1)$ and
$(2,2)$ inversion line observations from the Green Bank Ammonia Survey
\citep{2017ApJ...843...63F}, \cite{2019ApJ...872..207A} show that the
non-thermal velocity dispersion increases suddenly at the boundaries
of this and several other cores in Ophiuchus. Turbulent acceleration
of grains and grain-grain collisions can be envisioned to lead to an
enhanced desorption of frozen molecules in the transition zone. The
process should affect all atoms and molecules residing in the CO-rich
outer layers of grain mantles, and this would explain why $\meth$ and
SO have similar distributions.

In this scenario, the asymmetric distributions of $\meth$ and SO
around H-MM1 would either indicate that turbulence is stronger on the
eastern side of the core or that dust grain surfaces are richer in CO
on that side. The latter suggestion is supported by the fact that the
$\meth$ peak coincides with the dust temperature minimum
(Appendix~\ref{sec:tdust}). The former condition cannot be properly tested
using the present data because the boundaries are not probed with any
other lines than methanol and SO; the $\dammo$ line emission is
confined to the inner regions where the lines are narrow. The angular
resolution of the Green Bank $\ammo$ maps ($32\arcsec$;
\citealt{2017ApJ...843...63F}), which cover both the core and the
ambient cloud, is not sufficient for detailed comparison of the
velocity dispersions at the eastern and western
boundaries. Nevertheless, a similar analysis as performed by
\cite{2019ApJ...872..207A}, but averaging over semicircles, shows that
the non-thermal velocity dispersion (measured along the line of sight)
grows more slowly on the eastern side than on the western side,
contradicting the supposition of a more vigorous turbulence on that
side. A diagram illustrating this difference is shown in
Figure~\ref{figure:sigmas}.

\subsection{Desorption in  a shear layer}

\label{sec:hmm1_shear}

\begin{figure}[htb]
\unitlength=1mm
\begin{picture}(80,54)(0,0)
\put(0,0){
\begin{picture}(0,0) 
\includegraphics[width=8.0cm,angle=0]{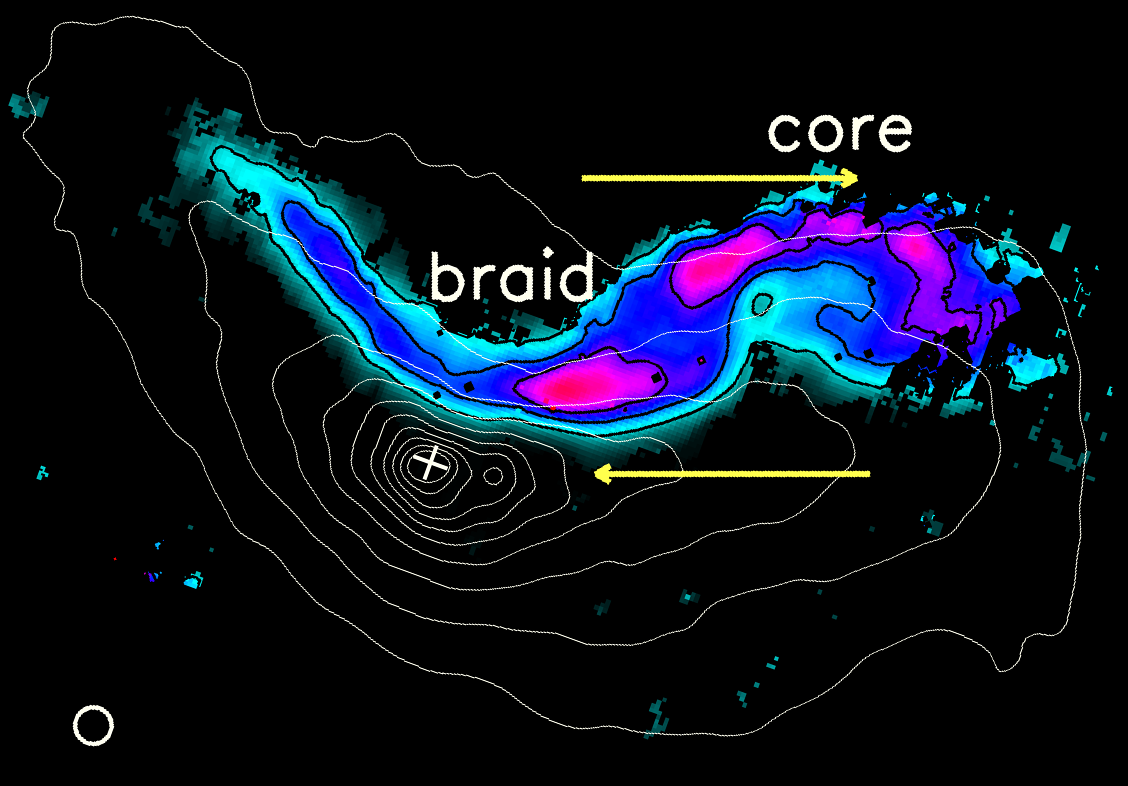} 
\end{picture}}
\end{picture}
\caption{Maps of the fractional $E-\meth$ abundance and the
  $\htwo$ column density (contours) already shown in Figure~\ref{figure:xmeth}
  but rotated here by $110\arcdeg$.  The arrows indicate the
  hypothesized flow directions in the denser and thinner gas components,
  relative to the average speed of the flow. The arrow length,
  $30\arcsec$, corresponds approximately one half the length of the
  wave-like structure.}
\label{figure:hmm1_khi_schematic}
\end{figure}

One remarkable feature of the brightest $\meth$ emission region is
that it closely follows the eastern boundary of the core as marked by
$\dammo$ emission.  The wavy shape and the rolling-up seen in the
north (Figure~\ref{figure:xmeth}) are the hallmarks of
Kelvin-Helmholtz instability (KHI), which can occur in sheared flows
with density stratification. The channel maps shown in
Figure~\ref{figure:channel_maps} give the impression that the dense
gas is flowing along a loop.  If the ambient gas does not share this
motion, the core is surrounded by a shear layer.  The presence of
shear in a gas flow with a high Reynolds number would imply
small-scale vorticity, and the undulating methanol emission region
would in this case consist of a chain of secondary billows that are
unresolved in the present map.  In this scenario, the methanol peak
can be identified with a ``braid'' with crowded isopycnic surfaces,
and the overturning billow in the north with a ``core'' where these
surfaces roll up (\citealt{1976JFM....73..215P}; see
Section~\ref{sec:shear}). The apparent wavelength of the methanol
feature is approximately $60\arcsec$, corresponding to $8,400$\,au or
$0.041$\,pc. The conceived structure is illustrated in
  Figure~\ref{figure:hmm1_khi_schematic}, which shows the maps of
  Figure~\ref{figure:xmeth} rotated in such a way that the suggested
  shear layer is approximately horizontal and the density increases
  downward. The directions of the relative velocities of the denser and
  thinner gas components are shown with arrows, and the positions of
  the ``braid'' and ``core'' are indicated.

In what follows, we attempt to estimate the flow properties, assuming
that, in analogy with atmospheric billow clouds, the largest
wavelength corresponds to the so called internal gravity waves, which
oscillate at the Brunt-V{\"a}is{\"a}l{\"a} frequency. This frequency,
also known as the buoyancy frequency, is defined by
$N=\sqrt{\frac{-g}{\rho}\frac{\partial{\rho}}{\partial{z}}}$, where
$g$ is the gravitational acceleration (directed to the negative $z$
direction) and $\rho$ is the gas density. We estimated the density and
the density gradient in the supposed shear layer by fitting a
Plummer-type function to the $\htwo$ column density profile from the IRAC
$8\,\mu$m absorption. The inversion method is explained by
\cite{2011A&A...529L...6A}.

The fit was made to the cross-sectional profile along an axis going
through the methanol peak at RA 16:27:59.5, Dec. -24:33:30
(J2000). The axis is tilted with respect to the R.A. axis by
$25^\circ$. According to this fit, the number densities at methanol
peak and at the spine of cloud are $\sim 3.8\times10^5\,{\rm cm}^{-3}$
and $\sim 9.2\times 10^5\,{\rm cm}^{-3}$, respectively.  The separation
between these points is approximately $10\arcsec$ (1,400\,au).  We
assumed that the gravitational field at the methanol peak is dominated
by the mass contained in a circular region of a radius of $10\arcsec$,
centered at the crossing of the cloud spine and the cross-sectional
axis. This mass is $0.11\,M_\odot$. The Brunt-V{\"a}is{\"a}l{\"a}
frequency obtained is $N \sim 1.3\times10^{-12}$\,Hz
(period 150,000 yr). The multiplication of $N$ by the apparent wavelength
gives a phase velocity of $\sim 260\,{\rm m\,s^{-1}}$. This should
correspond to the average speed of the sheared flow.

In case the flow is subject to KHI, a condition is imposed to
the velocity shear: the Richardson number, defined by ${\rm Ri} =
\frac{N^2}{(\partial{v}/\partial{z})^2}$, is less than $1/4$. The
implied minimum shear is $\sim 2.6\times 10^{-12}\,{\rm s}^{-1}$ or
$\sim 80\, \kms\,{\rm pc}^{-1}$. This value exceeds the north-south
velocity gradient derived from the molecular line maps by a factor of
4. Assuming that the velocity at the outer boundary of the shear layer
does not exceed $400\,{\rm m\,s^{-1}}$, which is the typical non-thermal
velocity far from the core, the maximum thickness of the layer is
approximately 1000\,au.  This estimate agrees with the
thickness of the methanol layer. 

In view of the uncertainties concerning the ``vertical'' scale and the
structure of the vorticity, we do not attempt to estimate methanol
production rate in the suggested vortex sheet. We merely state that
the present observations indicate that the formation of methanol is more
efficient in this layer than elsewhere in the core envelope. We
suggest, based on the theoretical and numerical studies quoted above,
that this is caused by the fact that two-dimensional turbulence can
have larger velocity fluctuations than full three-dimensional
turbulence on a time scale comparable to the drag time of dust grains.

\section{Conclusions}

\label{sec:concl}

Gaseous methanol and sulphur monoxide coexist on the outskirts of the
prestellar core H-MM1 in Ophiuchus. They are confined to a halo which
follows the core boundaries and which is particularly prominent on the
eastern side of the core.  Because methanol is mainly produced on
grains, the emission indicates regions of enhanced desorption. Sulphur
monoxide may have co-desorbed with $\meth$ or formed in the gas phase
following the release of other S-bearing species. The
  distribution of deuterated ammonia, $\dammo$, follows the $\htwo$
  column density distribution, and is almost complementary to those of
  $\meth$ and SO. This can be understood by the fact that deuterium
  fractionation is most efficient in the gas component where CO and
  other heavier species are strongly depleted, which is equivalent to
  ineffective desorption.

The asymmetry of the $\meth$ and SO distributions is partly caused 
by the strong, uneven illumination of the core,
  which mainly comes from luminous B-type stars on its western side. Our
  radiative transfer calculations indicate that the FUV field, albeit
  heavily attenuated, is significantly stronger on the western boundary
  of the core, with weak $\meth$ and SO emission, than on the eastern
  boundary, where these molecules emit strongly. The asymmetric FUV
  field has two effects: Firstly, the photodissociation rate of $\meth$ is
  higher on the western side. Secondly, the dust temperature has a minimum
  on the eastern side of the core. The minimum coincides with 
  the methanol maximum. This is understandable as CO accretion and 
hydrogenation are favored at low temperatures.

We suggest that external radiation is not the sole factor causing
  the contrast between the $\meth$ and SO abundances on the two sides
  of the core, but that this contrast is increased by forceful
  desorption on its shaded, eastern boundary. We consider it possible
  that grain-grain collisions make a significant contribution to
  desorption, along with cosmic ray collisions and spontaneous
  exothermic surface reactions. Desorption by grain-grain collisions
  can proceed through mild heating that triggers explosive
  radical-radical reactions and partial disruption of grain mantles
  \citep{1982A&A...109L..12D}, or through direct heating in energetic
  collisions. Collisions between grains can be induced by Kolmogorov
turbulence (\citealt{1985prpl.conf..621D};
\citealt{2002ApJ...566L.105L}), or by shear vorticity.

The brightest methanol emission region at the eastern boundary of the
core shows signatures of Kelvin-Helmholtz instability, indicating
strong velocity shear. On the other hand, the non-thermal velocity
dispersion along the line of sight, as traced by $\ammo$ lines, grows
more steeply on the opposite side of the core, where both $\meth$ and
SO emissions are weak. We interpret this so that at the eastern
boundary, laminar flow occurring mainly in the plane of the sky
is currently transitioning into turbulence through shear instability,
whereas on the western side, the flow has already developed into full
three-dimensional turbulence. The asymmetries of the $\meth$ and SO
distributions suggest that shear vorticity induces more energetic or
more frequent grain-grain collisions than Kolmogorov turbulence. This
is likely to be related to the fact that two-dimensional turbulence
has a steeper energy spectrum \citep{1971JFM....47..525K}, implying
larger velocity fluctuations on the time scale needed to accelerate
grains through hydrodynamic drag.

\vspace{2mm}

We thank Tom Hartquist, David Williams, Kalevi Mattila, and Hannu
Savij{\"a}rvi for helpful and enjoyable conversations, and the anonymous referee for critical comments that helped to improve this paper.  The paper makes
use of the following ALMA data: ADS/JAO.ALMA\#2016.1.00035.S. ALMA is
a partnership of ESO (representing its member states), NSF (USA) and
NINS (Japan), together with NRC (Canada), MOST and ASIAA (Taiwan), and
KASI (Republic of Korea), in cooperation with the Republic of
Chile. The Joint ALMA Observatory is operated by ESO, AUI/NRAO and
NAOJ.  The National Radio Astronomy Observatory is a facility of the
National Science Foundation operated under cooperative agreement by
Associated Universities, Inc.

This work is based in part on observations made with the Spitzer Space
Telescope, and made use of the NASA/IPAC Infrared Science Archive.
These facilities are operated by the Jet Propulsion Laboratory,
California Institute of Technology under a contract with National
Aeronautics and Space Administration. This work was supported by the
Academy of Finland (grants 285769 and 307157). Work by AV and AP is
supported by Russian Science Foundation via the Project
18-12-00351. AV and AP are members of the Max Planck Partner Group at
the Ural Federal University.

\bibliographystyle{aasjournal} 

\bibliography{AAS23476R1}

\appendix

\section{Column density map from 8\,$\mum$ extinction}

\label{sec:column}
 
The core is largely obscured from ultraviolet radiation and it
serves as an absorbing component at 8\,$\mum$. The surface brightness
in the core region can therefore be written as $I^{8\,\mum} = I_{\rm
  bg}^{8\,\mum}\,e^{-\tau^{8\,\mum}} + I_{\rm fg}^{8\,\mum}$, where
$\tau^{8\,\mum}$ is the optical thickness at 8\,$\mum$, $I_{\rm bg}$
is the surface brightness at the background of the core, and $I_{\rm
  fg}$ is the surface brightness of the foreground component (see,
e.g., \citealt{2009ApJ...696..484B}). The latter term may also contain
a zero point offset. The optical thickness map can be obtained from
$$
\tau^{8\,\mum} = -\ln \left \{ \frac{I^{8\,\mum}-I_{\rm fg}^{8\,\mum}}
{ I_{\rm bg}^{8\,\mum}} \right \} \; .
$$
In order to estimate the background emission, $I_{\rm bg}^{8\,\mum}$,
we made a fit to the surface brightness distribution with the core
masked out.  For masking we used the SCUBA-2 $850\,\mum$ map. The
foreground component, $I_{\rm fg}^{8\,\mum}$ (including possible
zero-point offset), was assumed to be constant in the mapped
region. Its value was estimated by requiring that the ratio of the
peak optical thicknesses is $\tau^{8\,\mum}/\tau^{850\,\mum}=780$,
corresponding to the adopted dust opacity model, which was the model
for unprocessed dust grains with thin ice mantles by
\cite{1994A&A...291..943O}. The comparison was done using an IRAC map 
smoothed to the resolutions of the SCUBA map. 

At $850\,\mum$, the core is optically thin and it is seen in thermal
dust emission. The observing method filters out extended emission. The
surface brightness is therefore $I^{850\,\mum} = B^{850\,\mum}(T_{\rm
  d})\,\tau^{850\,\mum}$, where $T_{\rm d}$ is the dust temperature
and $B$ is the Planck function.  The opacity map at $850\,\mum$,
$\tau^{850\,\mum}$, was calculated by combining the $850\,\mum$
surface brightness map with the dust color temperature map, $T_{\rm
  C}$, derived from Herschel/SPIRE maps
\citep{2017A&A...600A..61H}. The angular resolution of the $T_{\rm C}$
map is $\sim 40\arcsec$, whereas for SCUBA it is approximately
$14\arcsec$. We think this discrepancy in the angular resolutions is
acceptable because the $850\,\mum$ surface brightness is not very
sensitive to the dust temperature (see below).

Finally, the $\htwo$ column density map shown in Fig.~\ref{figure:line_maps}d)
was calculated by dividing the $8\,\mum$ optical thickness map by the
absorption cross-section of dust per $\htwo$ molecule, which according
to the adopted dust model is $4.1\times10^{-23}\,{\rm cm^2\, \htwo\,
  molecule^{-1}}$ ($8.86\,{\rm cm^{2}\,g^{-1}}$).

\section{Dust temperature distribution}

\label{sec:tdust}

\begin{figure*}[]
\unitlength=1mm
\begin{picture}(160,80)(0,0)
\put(10,0){
\begin{picture}(0,0) 
\includegraphics[width=16.0cm,angle=0]{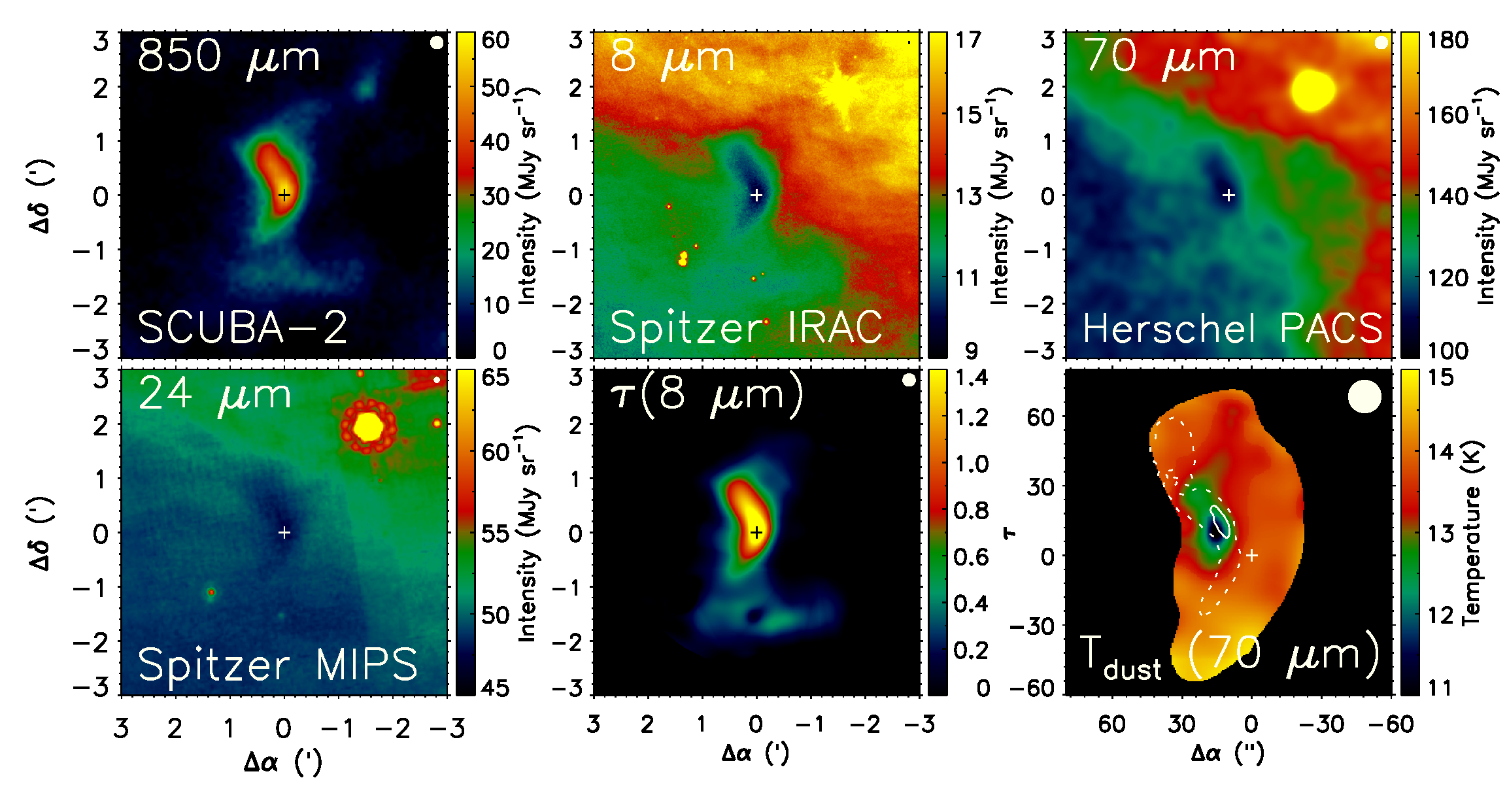} 
\end{picture}}
\put(28,74){\makebox(0,0){\large \bf \color{white} a)}}
\put(28,38){\makebox(0,0){\large \bf \color{white} b)}}
\put(78,74){\makebox(0,0){\large \bf \color{white} c)}}
\put(78,38){\makebox(0,0){\large \bf \color{white} d)}}
\put(128,74){\makebox(0,0){\large \bf \color{white} e)}}
\put(128,38){\makebox(0,0){\large \bf \color{white} f)}}
\end{picture}

\caption{Surface brightness maps of H-MM1 at selected mid- to
  far-infrared wavelengths (panels a to d), and the optical thickness
  (e) and the dust temperature (f) maps derived from these. The
  intruments and wavelengths are indicated in panels a) to d). The
  optical thickness map in panel d), $\tau^{8\,\mum}$, is
  derived from the Spitzer/IRAC map at $8\,\mum$. The dust temperature
  map, $T_{\rm dust}$, in panel f) is derived using the Herschel/PACS map
  at $70\,\mum$. Note that in this method, $T_{\rm dust}$ can be only
  derived when $\tau^{70\,\mum}$ is finite, and the region shown is
  smaller than in other panels. The maps presented in panels d), e)
  and f) are smoothed to the angular resolution of the SCUBA map in
  panel a) ($14\arcsec$).  The maps in panels b) and c) are shown at
  the original resolutions, $6\arcsec$ and $2\arcsec$,
  respectively. The angular resolution is indicated with a filled
  circle in the top right. The region of bright $\meth$ emission is
  outlined with dashed and solid contours in panel f). The plus sign
  indicates the position of the $\htwo$ column density maximum.}
\label{figure:taumidfar}
\end{figure*}
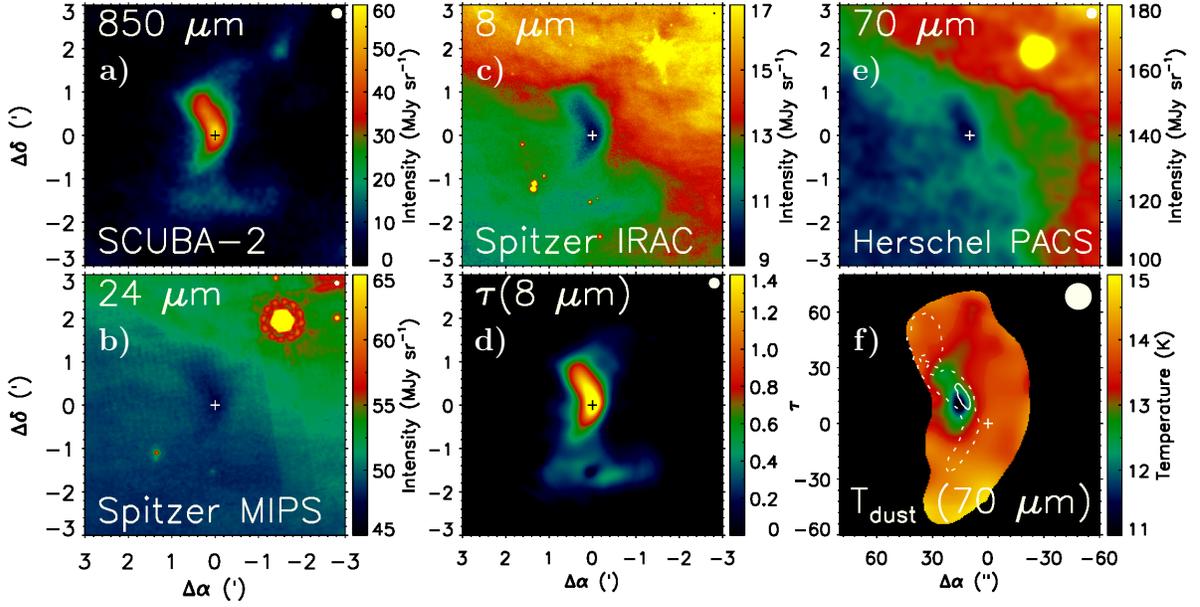

At $70\,\mum$, a cloud can be seen either as deficiency or excess
  of surface brightness depending on the dust temperature and the
brightness of the background.  At this wavelength, the Planck function
is very sensitive to small changes in the temperature. For example, a
temperature change from 12 to 14\,K for a given dust column increases
the emission by an order of magnitude, while at $850\,\mum$, the
corresponding change is $\sim 20\%$. The surface brightness at
$70\,\mum$ takes the form
$$
I^{70\,\mum} = I_{\rm bg}^{70\,\mum}\,e^{-\tau^{70\,\mum}} 
+ I_{\rm fg}^{70\,\mum}  + B^{70\,\mum}(T_{\rm d})(1-e^{-\tau^{70\,\mum}}) \; .
$$
We assumed that the $\tau^{70\,\mum}$ distribution is identical to
that at $8\,\mum$, save a constant factor depending on the adopted dust
opacity model for large ``classical'' grains (giving
$\tau^{8\,\mum}/\tau^{70\,\mum}=1.2$). The foreground level (including
a possible zero-point offset) was adjusted so that the dust
temperatures on the outskirts of the core are similar to $T_{\rm C}$
values derived from Herschel/SPIRE.  The obtained dust temperature
distribution in the core region is shown in the bottom right panel of
Figure~\ref{figure:taumidfar}. For the $T_{\rm d}$ calculation, the
$8\,\mum $ and $70\,\mum$ maps were smoothed to the $14\arcsec$
resolution of the SCUBA-2 map.

This map indicates that the dust temperature minimum is located
  on the eastern side of the column density maximum, and is nearly
coincident with the $\meth$ maximum.  We tested this result by
constructing a core model resembling H-MM1. In the plane of the sky,
the cloud structure was taken from the observed 850\,$\mu$m optical
depth map, and the line-of-sight density distribution has the same
shape as the $\tau^{850\,\mum}$ profile along horisontal cuts across
the core. The core was illuminated by an isotropic radiation field
plus a point source located on its western side. The dust opacity
model was the same as used above. The calculations were done using the
dust continuum radiative transfer program described in
\cite{2019A&A...622A..79J}. The isotropic field was set equal to
  30 times the \cite{1983A&A...128..212M} field, and the luminosity of
  point-source (modeled as a 10000\,K blackbody) was selected so that
  its contribution to the radiation field at the cloud center was
  equal to that of the isotropic background. The combined strength of
  the isotropic component and the point source were then adjusted
  until the $850\,\mum$ surface brightness map agreed with the
  observations.

\begin{figure}
\centering
\includegraphics[width=8.8cm]{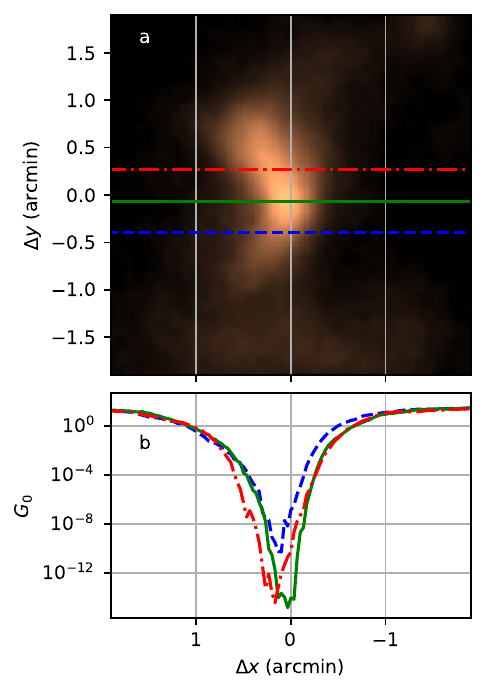}
\caption{
Strength of the UV radiation field inside the model cloud. Frame a
shows the locations of three horizontal cuts, drawn on a map of the
850\,$\mu$m optical depth. Frame b shows the 6.0-13.6\,eV energy
density along those lines, relative to the Habing field.
}
\label{figure:radiation_field}
\end{figure}

The dust temperature distribution derived from the simulated
$70\,\mum$ map in the same manner as described above shows a similar
shift of the minimum as seen in Figure~\ref{figure:taumidfar}f. In the
``true'', three-dimensional distribution, the dust temperature minimum
is shifted to the same direction, but lies a little closer to the
density peak. This experiment shows that dust temperatures derived
from the $70\,\mum$ surface brightness temperature map can correctly
reflect, although exaggerating slightly, the displacement of the dust
temperature minimum from the density maximum (and the $850\,\mum$
peak) caused by anisotropic illumination. 
Figure~\ref{figure:radiation_field} shows the energy density of the
6.0-13.6\,eV UV radiation field for one-dimensional cuts through the
densest part of the model. The energy density is plotted as $G_0$,
relative to the Habing value of $5.29\times 10^{-14}\,$erg\,cm$^{-3}$
\citep{1979ARA&A..17..345H}.

\section{Velocity dispersion profiles}

\label{sec:sigmav}

\begin{figure*}[]
\unitlength=1mm
\begin{picture}(160,130)(0,0)
\put(5,0){
\begin{picture}(0,0) 
\includegraphics[width=16.0cm,angle=0]{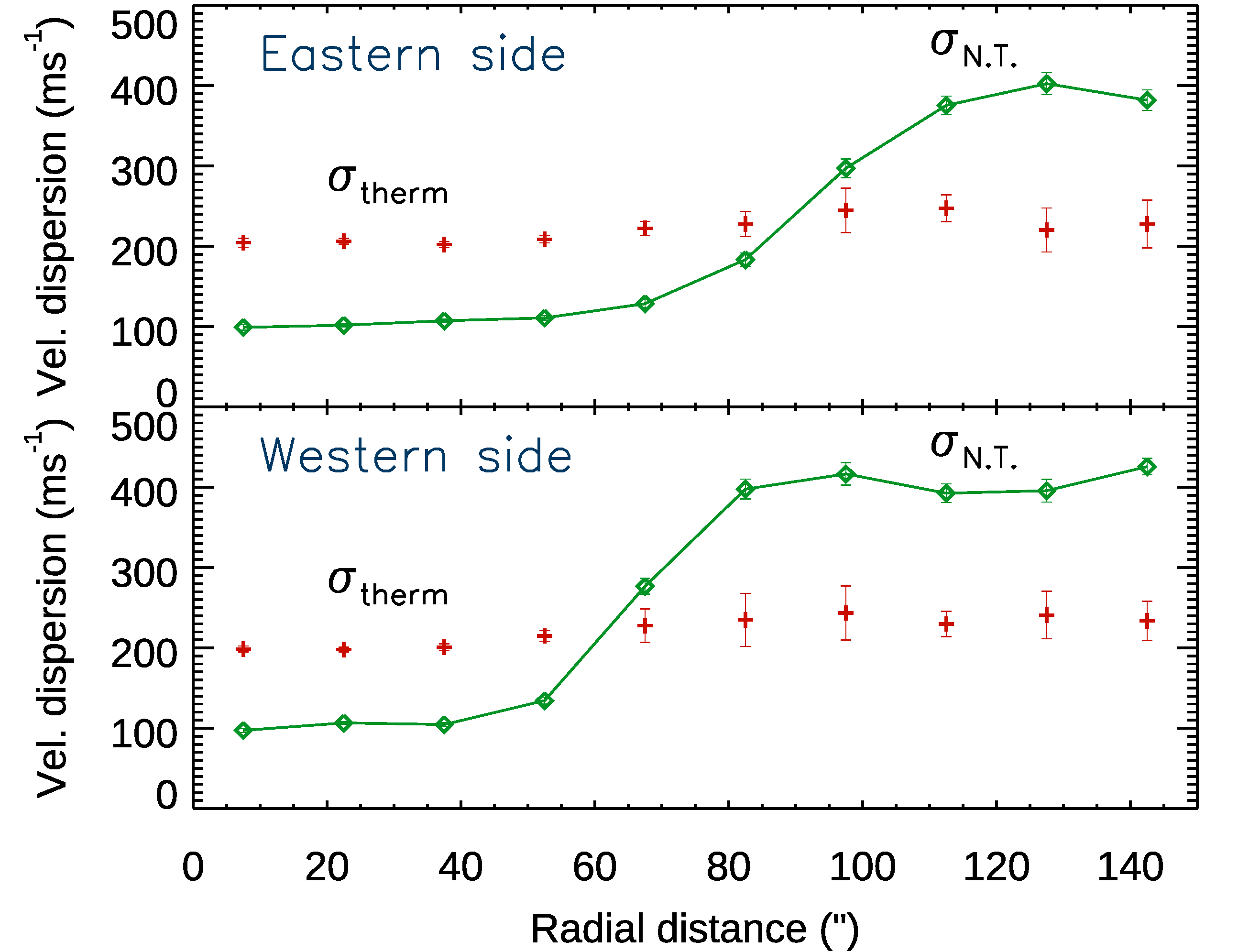} 
\end{picture}} 
\end{picture} 
\caption{Thermal and non-thermal velocity dispersions as functions of the
radial distance from the center on the eastern and western sides of H-MM1}
\label{figure:sigmas}
\end{figure*}

Thermal and non-thermal velocity dispersion profiles on the eastern
and western sides of the core were calculated using $\ammo(1,1)$ and
$(2,2)$ inversion line data from the Green Bank Ammonia Survey (GAS;
\citealt{2017ApJ...843...63F}). The method is described in detail by
\cite{2019ApJ...872..207A}. The spectra were first aligned in
  velocity using the Green Bank pipeline-reduced LSR velocity maps,
and then averaged in concentric semiannular regions. The stacked
$\ammo(1,1)$ and $(2,2)$ spectra were analyzed using the
standard method described, for example, by
\cite{1983ARA&A..21..239H}. This analysis gives estimates for the
kinetic temperature and total velocity dispersion (along the line of
sight). The radial distributions of thermal and non-thermal velocity
dispersions were calculated from these data. The results for the two
hemispheres of H-MM1 are shown in Figure~\ref{figure:sigmas}. The
non-thermal dispersion is approximately half the sound speed near the
center, and reaches a value of $\sim 400\,{\rm m\,s^{-1}}$, that is,
twice the sound speed far from the center of the core. The transition
from subsonic to supersonic regimes occurs, however, closer to the
center on the western side of the core than the eastern side, where
the strongest methanol emission comes from.

\end{document}